\documentclass[twocolumn,showpacs,preprintnumbers,amsmath,amssymb]{revtex4}
\usepackage{bm}
\newcommand{\eps}{\varepsilon}

\catcode`\@=11
\def\eqlt{\mathrel{\mathpalette\@vereq<}}  % < over =
\def\eqgt{\mathrel{\mathpalette\@vereq>}}  % > over =
\def\@vereq#1#2{\lower2.5pt\vbox{\baselineskip0pt \lineskip-.5pt
 \ialign{$\m@th#1\hfil##\hfil$\crcr#2\crcr{=}\crcr}}}

\usepackage{graphicx}% Include figure files
\begin{document}
%\draft
\title {Microscopic Model and Phase Diagrams of the Multiferroic Perovskite Manganites}
\author {Masahito Mochizuki$^{1}$ and Nobuo Furukawa$^{1,2}$}
\address {$^1$Multiferroics Project, ERATO, Japan Science and Technology Agency (JST),\\
c/o Department of Applied Physics, The University of Tokyo,\\
 7-3-1 Hongo, Bunkyo-ku, Tokyo 113-8656, Japan\\
$^2$Department of Physics and Mathematics, Aoyama Gakuin University, Sagamihara, Kanagawa 229-8558, Japan}
\date {today}

\begin{abstract}
Orthorhombically distorted perovskite manganites, $R$MnO$_3$ with $R$ being a trivalent rare-earth ion, exhibit a variety of magnetic and electric phases including multiferroic (i.e. concurrently magnetic and ferroelectric) phases and fascinating magnetoelectric phenomena. We theoretically study the phase diagram of $R$MnO$_3$ by constructing a microscopic spin model, which includes not only the superexchange interaction but also the single-ion anisotropy (SIA) and the Dzyaloshinsky-Moriya interaction (DMI). Analysis of this model using the Monte-Carlo method reproduces the experimental phase diagrams as functions of the $R$-ion radius, which contain two different multiferroic states, i.e. the $ab$-plane spin cycloid with ferroelectric polarization $P$$\parallel$$a$ and the $bc$-plane spin cycloid with $P$$\parallel$$c$. The orthorhombic lattice distortion or the second-neighbor spin exchanges enhanced by this distortion exquisitely controls the keen competition between these two phases through tuning the SIA and DMI energies. This leads to a lattice-distortion-induced reorientation of $P$ from $a$ to $c$ in agreement with the experiments. We also discuss spin structures in the A-type antiferromagnetic state, those in the cycloidal spin states, origin and nature of the sinusoidal collinear spin state, and many other issues.
\end{abstract}

%% 75.80.+q Magnetomechanical and magnetoelectric effects, magnetostriction
%% 77.80.-e Ferroelectricity and antiferroelectricity
%% 77.80.Bh Phase transitions and Curie point
%% 77.80.Dj Domain structure; hysteresis
%% 77.80.Fm Switching phenomena  
%% 75.30.-m Intrinsic properties of magnetically ordered materials
%% 75.30.Gw Magnetic anisotropy  
%% 75.47.-m Magnetotransport phenomena; materials for magnetotransport
%% 75.47.Lx Manganites 
\maketitle
%\sloppy \maketitle

\section{Introduction}
\label{Sec:Intro}
\begin{figure*}[tdp]
\includegraphics[scale=1.0]{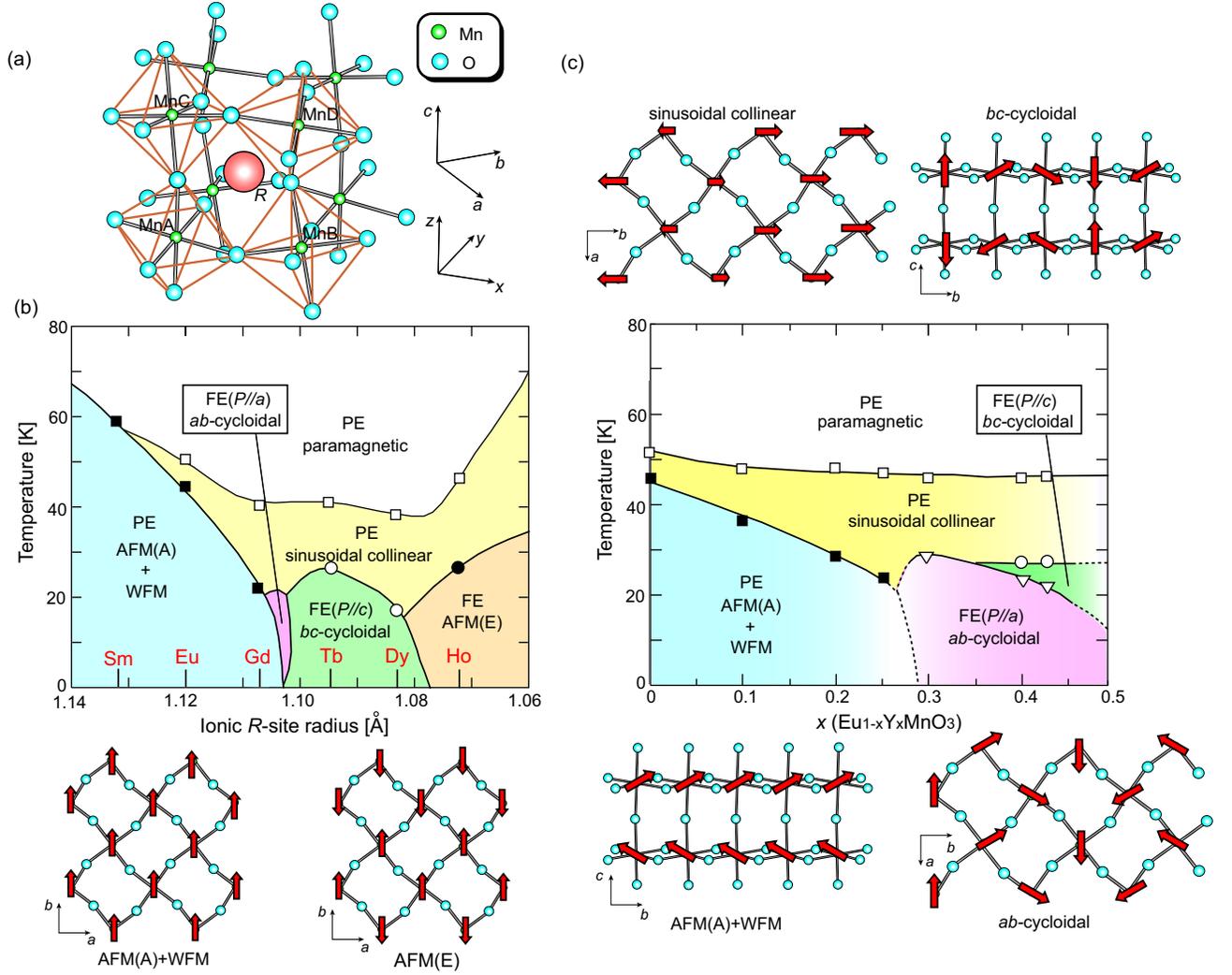}
\caption{(Color online) (a) Perovskite structure with GdFeO$_3$-type distortion. The unit cell contains four Mn ions, which are referred to as Mn A, Mn B, Mn C, and Mn D, respectively. (b) Experimentally obtained magnetoelectric phase diagram of $R$MnO$_3$ in plane of temperature and ionic $R$-site radius reproduced from Refs.~\cite{Kimura03b,Goto05}. The region between Gd and Tb was studied using a solid solution system Gd$_{1-x}$Tb$_x$MnO$_3$~\cite{Goto05}. Averaged $R$-site radii for the solid solutions are deduced by interpolation. Insets show spin configurations of the A-type and E-type antiferromagnetic [AFM(A) and AFM(E)] states. Along the $c$ axis, spins stack antiferromagnetically. PE and FE denote paraelectric and ferroelectric phases, respectively. In the AFM(E) phase, possible ferroelectricity due to an exchange-striction mechanism was theoretically proposed~\cite{Sergienko06b,Picozzi07,Yamauchi08}, and it was confirmed experimentally~\cite{Lorenz07,Lorenz04,PomjakushinCD09}. (c) Experimentally obtained magnetoelectric phase diagram of a solid solution system Eu$_{1-x}$Y$_x$MnO$_3$ in plane of temperature and Y concentration $x$ reproduced from Ref.~\cite{Yamasaki07b}. Insets show spin configurations of the $ab$-cycloidal, $bc$-cycloidal, and sinusoidal collinear states. Spin configuration of the AFM(A) state with weak ferromagnetism (WFM) seen along the $a$ axis is also shown.}
\label{Fig01}
\end{figure*}

Rare-earth manganites with orthorhombically distorted perovskite structure, $R$MnO$_3$ with $R$ being a trivalent rare-earth ion, have been subject to intensive studies since the multiferroic phases, in which magnetism and ferroelectricity simultaneously emerge, were found in some of these materials~\cite{Kimura03a,ReviewMF}. This class of materials exhibits a variety of magnetic and electric phases as a function of ionic radius of the $R$ ion. The $R$-site variation controls magnitude of the orthorhombic lattice distortion (the GdFeO$_3$-type distortion) ----- see Fig.~\ref{Fig01}(a) -----, i.e. tilting angles of the MnO$_6$ octahedra become larger and the perovskite lattice is more significantly distorted with a smaller-sized $R$ ion. An experimentally obtained phase diagram in Refs.~\cite{Kimura03b,JSZhou06} exhibits A-type antiferromagnetic [AFM(A)] ground states in the weakly distorted materials with $R$=La, Pr,..., Eu, Gd, while E-type antiferromagnetic [AFM(E)] ground states in the strongly distorted materials with $R$=Ho,..., Yb, Lu ----- see Fig.~\ref{Fig01}(b). Sandwiched by these two regions, a spiral spin order with concomitant ferroelectricity is observed in the moderately distorted materials with $R$=Tb and Dy.
%%These magnetic spirals are currently attracting appreciable interest 
%%as an origin of the multiferroic order, in which magnetism and 
%%ferroelectricity simultaneously emerge. 
Anticipated coupling between spins and electric dipoles (magnetoelectric coupling) makes these materials interesting because of possible technical applications.
%% for storage devices based on the electric control of magnetism 
%%and/or magnetic control of ferroelectricity.

%%%%%%%%%%%%%%%%%%%%%%%%%%%%%%%%%%%%%%%%%%%%%%%%%%%%%%%%%%%%%
%%%%%%%%%%%%%%%%%%%%%%%%%%%%%%%%%%%%%%%%%%%%%%%%%%%%%%%%%%%%%
In TbMnO$_3$, for example, a sinusoidal collinear order of Mn spins occurs at the N\'eel temperature $T_{\rm N}^{\rm Mn}\sim$~41 K~\cite{Quezel77,Kajimoto04}. In this phase, the Mn spins are aligned along the $b$ axis with an incommensurate propagation wave vector $\bm q_m^{\rm Mn}=$(0, 0.28, 1) in the $P_{bnm}$ orthorhombic unit cell. Below $T_{\rm C}\sim$~28 K, ferroelectricity shows up along the $c$ axis concomitantly with a magnetic transition into a transverse spiral (cycloidal) spin order with Mn spins rotating within the $bc$ plane~\cite{Kenzelmann05}. Upon further decreasing temperature, ordering of the $f$-electron moments on the rare-earth (Tb) ions takes place at $T_{\rm N}^R\sim$~7 K with a different propagation wave vector $\bm q_m^R\sim$(0, 0.42, 1). DyMnO$_3$ shows similar orderings and transitions with $T_{\rm N}^{\rm Mn}\sim$39 K, $\bm q_m^{\rm Mn}\sim$(0, 0.36, 1), $T_{\rm C}\sim$19 K, $T_{\rm N}^R\sim$~5 K, and $\bm q_m^R\sim$(0, 0.5, 1)~\cite{Feyerherm06}.

%%%%%%%%%%%%%%%%%%%%%%%%%%%%%%%%%%%%%%%%%%%%%%%%%%%%%%%%%%%%%
\begin{figure}[tdp]
\includegraphics[scale=1.0]{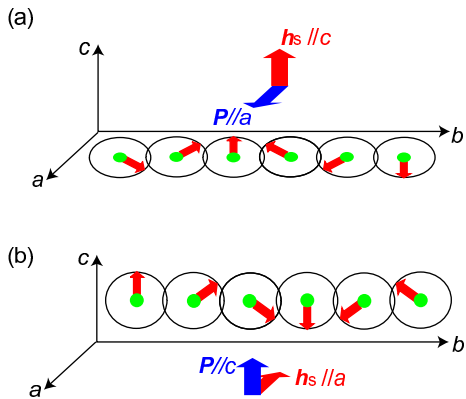}
\caption{(Color online) (a) Relationship between the spin-helicity vector $\bm h_s=\sum_{i}\bm S_i \times \bm S_{i+1}$ and the spontaneous electric polarization $\bm P$ in the $ab$-cycloidal spin structure expected from the spin-current model~\cite{Katsura05,Sergienko06a,Mostovoy06}, and (b) that in the $bc$-cycloidal spin structure.}
\label{Fig02}
\end{figure}
%%%%%%%%%%%%%%%%%%%%%%%%%%%%%%%%%%%%%%%%%%%%%%%%%%%%%%%%%%%%%
The ferroelectricity in these materials can be microscopically explained by the spin-current model [in other words, inverse Dzyaloshinsky-Moriya (DM) mechanism] proposed by Katsura, Nagaosa and Balatsky (KNB)~\cite{Katsura05}. Overlaps of electronic wave functions between adjacent two atomic sites ($i$ and $i+1$) with mutually canted spins ($\bm S_i$ and $\bm S_{i+1}$) can generate a local electric polarization,
%%%%%%%%%%%%%%%%%%%%%%%%%%%%%%%%%%%%%%%%%%%%%%%%%%
\begin{equation}
\bm p_i=A \bm e_{i,i+1} \times (\bm S_i \times \bm S_{i+1}),
\label{eqn:invDMmec}
\end{equation}
%%%%%%%%%%%%%%%%%%%%%%%%%%%%%%%%%%%%%%%%%%%%%%%%%%
where $\bm e_{i,i+1}$ denotes the unit vector connecting these two sites, and $A$ is a constant determined by the spin-exchange and spin-orbit interactions. The cycloidal spin order in $R$MnO$_3$ is expected to induce a uniform spontaneous polarization $\bm P$ as a sum of the local polarizations $\bm p_i$ in the direction perpendicular to the spiral propagation vector ($\parallel$$b$) and the spin-helicity vector, $\bm h_s=\sum_{i} \bm S_i \times \bm S_{i+1}$. Consequently, the ferroelectric polarization parallel to the $a$ axis ($P_a$) is expected in the $ab$-plane cycloidal spin structure, while in the $bc$-plane cycloidal spin structure, that parallel to the $c$ axis ($P_c$) is expected ----- see Fig.~\ref{Fig02}. Direct evidence for the inverse DM mechanism was recently demonstrated by controlling the spin-helicity vector in TbMnO$_3$ with an external electric field~\cite{Yamasaki07a}. Similar theoretical results have also been obtained independently in Refs.~\cite{Sergienko06a,Mostovoy06}. In particular, Sergienko and Dagotto discussed that a spiral spin order can induce uniform shifts of the oxygen ions through the spin-lattice coupling via the DM interactions, leading to the electric polarizations given by Eq.~(\ref{eqn:invDMmec}). Recent first-principles calculations revealed that the Sergienko-Dagotto type lattice-mediated contribution is dominant for the ferroelectric polarization relative to the KNB type pure-electronic contibution~\cite{HJXiang08,Malashevich08}.

While only the $P_c$ phase with $bc$-cycloidal spin structure is observed in the single-rare-earth compounds like TbMnO$_3$ and DyMnO$_3$, the $P_a$ phase with $ab$-cycloidal spin structure can be observed on the verge of the AFM(A) phase when we continuously control magnitude of the GdFeO$_3$-type distortion using a solid solution system Gd$_{1-x}$Tb$_x$MnO$_3$~\cite{Goto05}. Moreover, reorientation of the polarization occurs from $P_a$ to $P_c$ with increasing Tb concentration $x$  ----- see Fig.~\ref{Fig01}(b). This phenomenon is ascribed to a spin-cycloidal-plane flop from the $ab$ plane to the $bc$ plane~\cite{Yamasaki08}.

The reorientation of the electric polarization is also observed in another solid solution system Eu$_{1-x}$Y$_x$MnO$_3$~\cite{Hemberger07,Yamasaki07b}. Phase diagrams of this system were experimentally studied for compositions $0\leq x \lesssim0.5$ as functions of temperature and Y concentration $x$~\cite{Hemberger07,Yamasaki07b,Ivanov06a,Ivanov06b} ----- see Fig.~\ref{Fig01}(c). According to these experiments, the ground state changes approximately at $x\sim$~0.2-0.3 from the canted AFM(A) state without long-range ferroelectric order towards the presumably $ab$-cycloidal spin state with $P_a$. For higher Y concentrations of $x\geq0.4$, the orientation of polarization spontaneously changes from $P_c$ at higher temperatures towards $P_a$ at lower temperatures. In addition, the regime of $P_c$ phase increases with increasing $x$, which resembles the polarization flop with increasing $x$ in the Gd$_{1-x}$Tb$_x$MnO$_3$ system. These results indicate that the polarization flop occurs both thermally and by the increase of GdFeO$_3$-type distortion. These phenomena are caused purely by interactions among the Mn $3d$ spins since the Eu$_{1-x}$Y$_x$MnO$_3$ system is free from the influence of $f$-electron moments on the $R$ ions.

The above experimental results tell us that the manganite system exhibits the AFM(A), $ab$-cycloidal, $bc$-cycloidal, and AFM(E) phases successively as the GdFeO$_3$-type distortion increases. In particular, the cycloidal-plane flop between $ab$ and $bc$ is quite important since the plane is related to the orientation of electric polarization. According to Ref.~\cite{Kimura03b}, enhanced tilting of the MnO$_6$ octahedra leads to an increase of the second-neighbor antiferromagnetic exchanges ($J_2$), which compete with the nearest-neighbor ferromagnetic exchanges ($J_1$) in the $ab$ plane. Resulting weakening of the effective in-plane ferromagnetic interactions causes reduction of the N\'eel temperature in the AFM(A) phase as the (averaged) $R$-site radius decreases. We can attribute the emergence of the long-wave-length antiferromagnetic orders or that of the spiral spin orders to magnetic frustration caused by the above competition. The spiral spin order can be reproduced by a simple $J_1$-$J_2$ Heisenberg model~\cite{Yoshimori59,Kaplan59,Villain59}. Within this model, however, there exists degeneracy and the cycloidal plane cannot be specified so that the observed cycloidal-plane flop cannot be explained.

Below we summarize puzzling issues in the experimental phase diagrams of $R$MnO$_3$:
\begin{itemize}
\item An issue how the $bc$-cycloidal order is stabilized in the strongly distorted materials despite the fact that in the perovskite manganites, the $c$ axis is always a hard magnetization axis. 
\item A mechanism of the spin-cycloidal-plane (electric-polarization) flop with increasing GdFeO$_3$-type distortion.
\item A mechanism of the thermally induced cycloidal-plane flop or an issue why the $bc$-cycloidal phase appears above the $ab$-cycloidal phase in temperature.
\item Origin and nature of the sinusoidal collinear phase in the intermediate temperature region.
\item An issue why the Mn spins in the AFM(A) and the sinusoidal collinear states direct along the orthorhombic $b$ axis~\cite{Wollan55,Wollan57,Elemans71,QuezelA68,Jirak85}.
\end{itemize}
These issues are interesting not only for potential applications but also for fundamental physics. 

Previously, Ishihara and coworkers theoretically studied the phase diagram with the use of a two-dimensional $J_1$-$J_2$ Ising model~\cite{Kimura03b}. Although they predicted long wave-length magnetic structures between AFM(A) and AFM(E) phases, they inevitably failed to reproduce the cycloidal spin order and the sinusoidal collinear spin order since they treated the Mn spins as Ising spins. Recently, Dagotto and coworkers theoretically studied the phase diagram and the cycloidal spin order by employing a two-orbital double-exchange model on the two-dimensional lattice with some additional terms~\cite{Sergienko06a,DongS08}. In Ref.~\cite{Sergienko06a}, they incorporated the lattice elastic energy as well as the DM interactions with vectors on the Mn-O-Mn bonds coupling to the oxygen displacements, but neglected the second-neighbor spin exchanges. They discussed that the cycloidal spin order is stabilized due to the ferri-type arrangement of DM vectors realized by uniform shifts of the oxygens. In Ref.~\cite{DongS08}, they considered quite weak next-neighbor spin exchanges, but neglected the DM interactions. They reproduced the successive emergence of three magnetic phases in the ground state, i.e., AFM(A), cycloidal, and AFM(E) phases. However, in those studies, neither the spin-cycloidal-plane flop nor the sinusoidal collinear state was reproduced. In addition, the issues on the easy-axis spin anisotropy in the AFM(A) phase and the emergence of $bc$-cycloidal spin state were not addressed. Therefore, all of the above-listed issues still remain unresolved.

In this paper, we theoretically investigate the magnetoelectric phase diagrams of $R$MnO$_3$ in the absence of external magnetic field. We construct a microscopic spin model, which is basically a classical $S$=2 Heisenberg model on a cubic lattice, but includes some additional magnetic anisotropies and interactions~\cite{Mochizuki09}. By analyzing this model using the Monte-Carlo method, we obtain phase diagrams in good agreement with experiments. We discuss a number of puzzling issues on the magnetoelectric phase diagrams of the perovskite manganites listed above by particularly focusing on the mechanism of the spin-cycloidal-plane flop. We reveal that the DM interaction between spins neighboring along the $x$ or $y$ axis as well as the single-ion anisotropies, because of which the $c$ axis becomes a hard magnetization axis, favor the $ab$-cycloidal spin state with $P_a$, while the DM interaction between spins neighboring along the $c$ axis favors the $bc$-cycloidal spin state with $P_c$. Their competition turns out to be controlled by the GdFeO$_3$-type distortion, or equivalently by the second-neighbor exchanges $J_2$ enhanced by this lattice distortion. This leads to a polarization flop from $P_a$ to $P_c$ with decreasing $R$-site radius in agreement with the experiments.

Here we note that although recent intensive experiments on the present manganites have uncovered rich phase diagrams and phenomena also in the presence of magnetic fields~\cite{Kimura03a,Yamasaki07b,Kimura05,Abe07,Murakawa08,Goto04,Kagawa09,Schrettle09,Arima05,Noda05,Noda06,Strempfer07,Aliouane06,Argyriou07,Meier07,Strempfer08,Kadomtseva05a,Kadomtseva05b,MTokunagaUP}, these issues are beyond the scope of this paper. However, a microscopic model constructed in this paper would offer a very powerful basis for studying the magnetic-field effects in the multiferroic manganites as well.

Moreover, several fascinating phenomena originating from the strong magnetoelectric coupling have been observed in the multiferroic phases in $R$MnO$_3$, and have attracted appreciable interest as listed below:
%%We list some of these phenomena below
\begin{itemize}
\item The polarization flop can be induced also by an external magnetic field. In TbMnO$_3$ and DyMnO$_3$, for instance, the electric polarization flops from $P_c$ to $P_a$ by applying a magnetic field along the $a$ or $b$ axis~\cite{Kimura03a,Kimura05}. On the other hand, the polarization in Eu$_{1-x}$Y$_x$MnO$_3$ with $x=$0.4-0.5 flops from $P_a$ to $P_c$ with a magnetic field along the $a$ axis~\cite{Yamasaki07b}. In addition, control of the electric polarization vector using the rotating magnetic field was demonstrated experimentally~\cite{Abe07,Murakawa08}.

\item There are experimental~\cite{Smolenski82,Pimenov06a} and theoretical~\cite{Katsura07,Chupis07} arguments on possible low-lying spin excitations activated by the electric-field component of the light (termed electromagnon). Far-infrared spectroscopy observed corresponding light absorptions in the ferroelectric phases in TbMnO$_3$~\cite{Pimenov06a,Takahashi08,Aguilar09,Pimenov09}, DyMnO$_3$~\cite{Kida08}, GdMnO$_3$~\cite{Pimenov06a,Pimenov06b}, Gd$_{1-x}$Tb$_x$MnO$_3$~\cite{Kida08b} and Eu$_{1-x}$Y$_x$MnO$_3$~\cite{Aguilar07,Pimenov08} at terahertz frequencies.

\item Giant magnetocapacitance was observed at a threshold magnetic field for polarization flop in DyMnO$_3$~\cite{Kimura03a,Kimura05,Goto04}, and its mechanism has attracted interest~\cite{Kagawa09,Schrettle09}. On the basis of the dielectric-dispersion measurements, this phenomenon was attributed to the electric-field-driven motion of the multiferroic domain wall between domains with orthogonal spin-cycloidal planes and concomitant orthogonal ferroelectric polarizations~\cite{Kagawa09}. It was revealed that the macroscopic motion of the multiferroic domain wall is possible with electric fields of practical magnitude, which enables the electric control of magnetic domains.
\end{itemize}
It is widely recognized that interplay between charge, orbital, spin and lattice degrees of freedom is a key to understanding of these phenomena, similarly to the colossal magetoresistance in the hole-doped $R$MnO$_3$~\cite{ReviewCMR}. Therefore, construction of a microscopic model would represent a very important basis to approach these issues. Upon construction of a model, whether the experimental phase diagrams can be reproduced or not shall be an efficient test for its validity.

Furthermore, new insights into the magnetoelectric coupling obtained here would be useful when we study the related phenomena not only for the present manganites but also for other newly-discovered multiferroic materials. The magnetically-induced ferroelectricity has recently been found in several transverse-spiral magnets~\cite{Kimura07,Arima07} such as Ni$_3$V$_2$O$_8$~\cite{Lawes05}, MnWO$_4$~\cite{Taniguchi06,Heyer06}, LiCu$_2$O$_2$~\cite{SPark07,Seki08}, LiCuVO$_4$~\cite{Naito07} and CuO~\cite{Kimura08}, and also in a transverse cone-spiral magnet CoCr$_2$O$_4$~\cite{Yamasaki06}. Furthermore, it was demonstrated that a conical spin order induced by an external magnetic field also generates the electric polarization in ZnCr$_2$Se$_4$~\cite{Murakawa08a} and Y-type hexaferrites~\cite{Kimura05b,Ishiwata08}. Under this circumstance, clarification of the essential physics behind the rich phase diagrams, and a microscopic description of the magnetoelectric systems in the perovskite manganites have attained increasing importance because the manganite system is the first discovered example, and hence a prototype of a series of these new multiferroic materials.

The organization of this paper is as follows. In Sec.~\ref{Sec:ModelMethod}, we construct a microscopic spin model to describe the Mn 3$d$-spin system in the perovskite manganites. In Sec.~\ref{Sec:Results}, we discuss the calculated results obtained by analyzing this model in the Monte-Carlo simulations. We study phase diagrams and properties of several phases in Sec.~\ref{SSec:PhsDgm}. We address an issue on the spin-cycloidal-plane flop (or the electric-polarization flop) in Sec.~\ref{SSec:PFlop}. In Sec.~\ref{SSec:Ellipticity}, we also investigate detailed ground-state spin structures of the cycloidal spin states by numerically solving the Landau-Lifshitz-Gilbert equation. On this basis, we discuss the experimentally claimed elliptical modulation of the spin cycloid~\cite{Yamasaki07a,Arima06}. In Sec.~\ref{SSec:SIAAFMA}, by examining effects of the single-ion anisotropies, we address an issue why spins in the AFM(A) state and those in the sinusoidal collinear state are aligned parallel to the $b$ axis. In Sec.~\ref{SSec:SinColl}, we discuss the nature and origin of the sinusoidal collinear spin phase at intermediate temperatures. Section~\ref{Sec:Conc-Disc} is devoted to conclusion and discussion. We also explain how we calculate values of parameters in our model in Appendices A, B and C.
%%A short report on a part of the content of this paper has already been 
%%published~\cite{Mochizuki09}. 
%%In this paper, we describe a comprehensive framework together with 
%%additional new results and more detailed discussions.

\section{Model and Method}
\label{Sec:ModelMethod}
\subsection{Model Hamiltonian}
\label{SSec:Model}
To describe the Mn 3$d$-spin system in $R$MnO$_3$, we employ a classical Heisenberg model with some additional interactions and magnetic anisotropies on a cubic lattice. In this Hamiltonian, we treat the Mn $S$=2 spins as classical vectors, i.e. ${\bm S}_i$=($\sqrt{S^2-S_c^2}\cos\theta_i$, $\sqrt{S^2-S_c^2}\sin\theta_i$, $S_c$) with respect to the $P_{bnm}$ $a$, $b$ and $c$ axes. The Hamiltonian consists of four contributions as
%%%%%%%%%%%%%%%%%%%%%%%%%%%%%%%%%%%%%%%%%%%%%%%%%%
\begin{equation}
\mathcal{H}=\mathcal{H}_{\rm ex}+\mathcal{H}_{\rm sia}
+\mathcal{H}_{\rm DM}+\mathcal{H}_{\rm cub},
\label{eq:model1}
\end{equation}
%%%%%%%%%%%%%%%%%%%%%%%%%%%%%%%%%%%%%%%%%%%%%%%%%%
with
%%%%%%%%%%%%%%%%%%%%%%%%%%%%%%%%%%%
\begin{eqnarray}
\mathcal{H}_{\rm ex}&=&
-J_{ab}\sum_{<i,j>}^{x,y}\bm S_i \cdot \bm S_j
+J_2   \sum_{<i,j>}^{b}\bm S_i \cdot \bm S_j \\ \nonumber
& &+J_c   \sum_{<i,j>}^{c}\bm S_i \cdot \bm S_j,\\
\mathcal{H}_{\rm sia}&=&
D\sum_{i}S_{\zeta i}^2
+E\sum_{i}(-1)^{i_x+i_y}(S_{\xi i}^2-S_{\eta i}^2),\\
\mathcal{H}_{\rm DM}&=&
\sum_{<i,j>}\bm d_{ij}^{\alpha}\cdot(\bm S_i \times \bm S_j),\\
\mathcal{H}_{\rm cub}&=&\frac{a}{S(S+1)}\sum_{i}(S_{xi}^4+S_{yi}^4+S_{zi}^4),
\end{eqnarray}
%%%%%%%%%%%%%%%%%%%%%%%%%%%%%%%%%%%
where $i_x$, $i_y$ and $i_z$ represent coordinates of $i$-th Mn ion with respect to the cubic $x$, $y$ and $z$ axes. In the following, we explain each of these terms in detail, including definitions of other indices and variables.

\subsection{Superexchange interactions}
\label{SSec:Superexchange}
%%%%%%%%%%%%%%%%%%%%%%%%%%%%%%%%%%%%%%%%%%%%%%%%%%%%%%%%%%%%
\begin{figure}[tdp]
\includegraphics[scale=1.0]{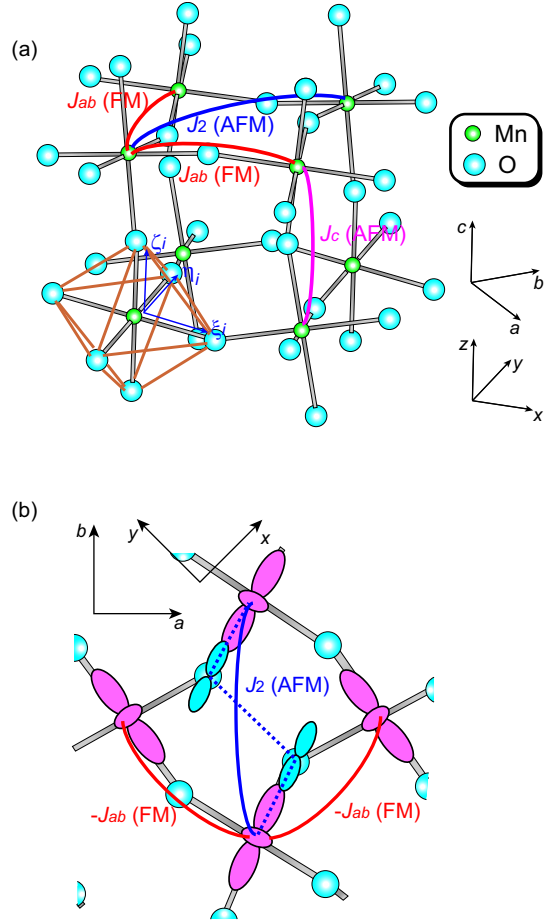}
\caption{(Color online) (a) Superexchange interactions in $R$MnO$_3$ and tilted local coordinate axes $\xi_i$, $\eta_i$, and $\zeta_i$ attached to the $i$-th MnO$_6$ octahedron. With regard to the superexchange interactions, we consider ferromagnetic exchanges $J_{ab}$ on the Mn-Mn bonds along the cubic $x$ and $y$ axes, antiferromagnetic exchanges $J_2$ on the in-plane diagonal Mn-Mn bonds along the orthorhombic $b$ axis, and antiferromagnetic exchanges $J_c$ on the Mn-Mn bonds along the $c$ axis. (b) Configuration of the occupied $e_g$ orbitals and superexchange interactions ($J_{ab}$ and $J_c$, solid curves) in the $ab$ plane. Exchange path for the superexchange $J_2$ via two oxygen $2p$ orbitals is indicated by a dashed line.}
\label{Fig03}
\end{figure}
%%%%%%%%%%%%%%%%%%%%%%%%%%%%%%%%%%%%%%%%%%%%%%%%%%%%%%%%%%%%%
The first term $\mathcal{H}_{\rm ex}$ represents superexchange interactions. This term contains ferromagnetic exchanges $J_{ab}$ on the Mn-Mn bonds along the $x$ and $y$ axes, antiferromagnetic exchanges $J_2$ on the in-plane diagonal Mn-Mn bonds along the orthorhombic $b$ axis, and antiferromagnetic exchanges $J_c$ on the Mn-Mn bonds along the $c$ axis ----- see Fig.~\ref{Fig03}(a). 

The Jahn-Teller distortion in $R$MnO$_3$ causes a C-type orbital ordering. Consequently, the ferromagnetic $J_{ab}$ is caused by the antiferro arrangement of $e_g$ orbitals in the $ab$ plane, while along the $c$ axis the antiferromagnetic $J_c$ is caused by the ferro-orbital stacking. 
%%On the other hand, the antiferromagnetic $J_2$ is caused by the 
%%finite overlap of the $e_g$ orbitals in the $b$-direction mediated 
%%by the two oxygens owing to the GdFeO$_3$-type distortion 
%%[see Fig.~\ref{Fig03}(b)].
On the other hand, the antiferromagnetic $J_2$ arises from an exchange path between two Mn $e_g$ orbitals in the $b$ direction via two oxygen $2p$ orbitals as shown by a dashed line in Fig.~\ref{Fig03}(b). The strength of the second-neighbor exchanges $J_2$ increases with increasing GdFeO$_3$-type distortion since the distortion enhances hybridization between these two oxygen $2p$ orbitals~\cite{Kimura03b,Picozzi06}.

%%%%%%%%%%%%%%%%%%%%%%%%%%%%%%%%%%%%%%%%%%%%%%%%%%%%%%%%%%%%
\begin{figure}[tdp]
\includegraphics[scale=1.0]{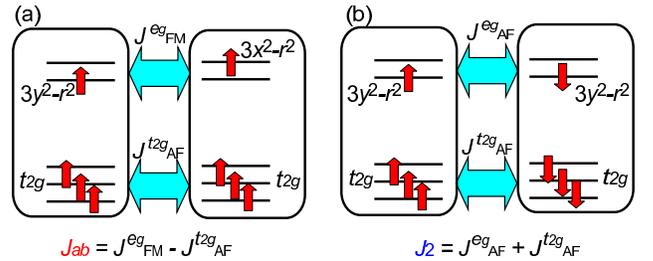}
\caption{(Color online) (a) Opposite contributions to the nearest-neighbor ferromagnetic exchange $J_{ab}$ from the $t_{2g}$- and $e_g$-orbital sectors. (b) Cooperative contributions to the second-neighbor antiferromagnetic exchange $J_2$ from the $t_{2g}$- and $e_g$-orbital sectors.}
\label{Fig04}
\end{figure}
%%%%%%%%%%%%%%%%%%%%%%%%%%%%%%%%%%%%%%%%%%%%%%%%%%%%%%%%%%%%%
It should be mentioned that in other orthorhombic perovskite compounds, the strength of the second-neighbor exchanges $J_2$ is negligibly weak as compared to that of the nearest-neighbor exchanges $J_{ab}$. By contrast, in the manganites, these two exchanges $J_2$ and $J_{ab}$ are comparable in strength, and the second-neighbor exchanges $J_2$ play important roles in determining the magnetic properties. This can be attributed to the electronic structure of Mn$^{3+}$ ion with $t_{2g}^3e_g^1$ electron configuration. For the nearest-neighbor Mn-Mn pairs, there are two opposite contributions to the superexchange $J_{ab}$ as shown in Fig.~\ref{Fig04}(a). The exchange between the $S$=1/2 spins on the staggered $e_g$ orbitals gives a ferromagnetic contribution, while in the $t_{2g}$-orbital sector, the antiferromagnetic coupling is realized between the $S$=3/2 spins. The cancellation of these opposite contributions results in the small magnitude of $J_{ab}$. On the other hand, for the Mn-Mn pairs neighboring along the $b$ axis, both the $e_g$-orbital and the $t_{2g}$-orbital sectors give antiferromagnetic contributions to the superexchange $J_2$ as in Fig.~\ref{Fig04}(b), resulting in the relatively large magnitude of $J_2$. The frustration between the ferromagnetic $J_{ab}$ and the antiferromagnetic $J_2$ stabilizes a spiral spin order in $R$MnO$_3$ compounds with relatively strong GdFeO$_3$-type distortion like TbMnO$_3$, DyMnO$_3$, and certain kinds of solid solutions.

%%%%%%%%%%%%%%%%%%%%%%%%%%%%%%%%%%%%%%%%%%%%%%%%%%%%%%%%%%%%%%%%%%
\begin{table}
\caption{Calculated superexchange parameters ($J_{ab}$ and $J_c$) and single-ion anisotropy parameters ($D$ and $E$) for several $R$MnO$_3$ compounds. Structural data used in the calculations are taken from Refs.~\cite{RodriguezC98,Alonso00,Mori02,Dabrowski05,Tachibana07}. In the second row, their ground-state magnetic structures are presented. Calculated values of $D$ and $E$ contain some ambiguities due to uncertainty in the optical absorption data~\cite{Matsumoto70,Gerristen63}. The values in the upper (lower) rows are calculated using the data from Ref.~\cite{Matsumoto70} (Ref.~\cite{Gerristen63}). Values of the second-neighbor antiferromagnetic exchange $J_2$ are also presented in the last row, which are calculated using the $J_1$-$J_2$ model. (see Apendix B) The value of $J_2$ in EuMnO$_3$ [AFM(A)] is deduced by extrapolation.}
\begin{tabular}{c|cccc}
\hline
 $R$MnO$_3$ & EuMnO$_3$ & TbMnO$_3$ & DyMnO$_3$ & YMnO$_3$ \\
\hline
Magnetic state & AFM(A) & $bc$-cycloidal & $bc$-cycloidal & cycloidal? \\
\hline
$J_{ab}$ (meV) & 0.85 & 0.79 & 0.76 & 0.77 \\
$J_c$ (meV)    & 1.27 & 1.26 & 1.20 & 1.25 \\
\hline
$D$ (meV)      & 0.32 & 0.33 & 0.33 & 0.34 \\
               & 0.24 & 0.25 & 0.25 & 0.25 \\
$E$ (meV)      & 0.34 & 0.34 & 0.34 & 0.34 \\
               & 0.25 & 0.24 & 0.24 & 0.24 \\
\hline
$J_2$ (meV)    & $\sim$0.5  & 0.62 & 0.92 & 1.9 \\
\hline
\end{tabular}
\label{tabl:PRMVL}
\end{table}
%%%%%%%%%%%%%%%%%%%%%%%%%%%%%%%%%%%%%%%%%%%%%%%%%%%%%%%%%%%%%%%%%%
We calculate the values of superexchange parameters $J_{ab}$ and $J_c$ for several $R$MnO$_3$ compounds by using formulae given in Refs.~\cite{Gontchar01,Gontchar02}. The results are listed in Table~\ref{tabl:PRMVL}. The structural data used in this calculation are taken from Refs.~\cite{RodriguezC98,Alonso00,Mori02,Dabrowski05,Tachibana07}. The changes of their values are very small upon the $R$-site variation as far as vicinities of the multiferroic phases are concerned. We take $J_{ab}$=0.80 meV and $J_c$=1.25 meV in the following calculations. We also estimate the values of $J_2$ in TbMnO$_3$, DyMnO$_3$, YMnO$_3$ and several solid solutions by employing the two-dimensional $J_1$-$J_2$ classical Heisenberg model, which reveals that the value of $J_2$ monotonically increases with decreasing $R$-site radius. For the details, see Appendices A and B.

\subsection{Single-ion anisotropy}
\label{SSec:SIAnisotropy}
%%%%%%%%%%%%%%%%%%%%%%%%%%%%%%%%%%%%%%%%%%%%%%%%%%%%%%%%%%%%%%%%%%
\begin{table}
\caption{Structural parameters of EuMnO$_3$ from Ref.~\cite{Dabrowski05}.}
\begin{tabular}{cccccccc}
\hline
 $a$ ($\AA$) & $b$ ($\AA$) & $c$ ($\AA$) & $x_{\rm O_1}$ & $y_{\rm O_1}$ & 
 $x_{\rm O_2}$ & $y_{\rm O_2}$ & $z_{\rm O_2}$ \\
\hline
5.3437 & 5.8361 & 7.46186 & 0.0974 & 0.4714 & 0.7055 & 0.3247&  0.04845 \\
\hline
\end{tabular}
\label{tabl:SDEMO}
\end{table}
%%%%%%%%%%%%%%%%%%%%%%%%%%%%%%%%%%%%%%%%%%%%%%%%%%%%%%%%%%%%%%%%%%
The second term $\mathcal{H}_{\rm sia}$ denotes the single-ion anisotropies, which is determined by the wave function of occupied $e_g$ orbital, or equivalently by the local environment of Mn$^{3+}$ ion surrounded by six oxygens. Here $\xi_i$, $\eta_i$ and $\zeta_i$ are tilted local axes attached to the $i$-th MnO$_6$ octahedron as shown Fig.~\ref{Fig03}(a). This term consists of two parts, i.e. $\mathcal{H}_{\rm sia}^D$ and $\mathcal{H}_{\rm sia}^E$. The former part $\mathcal{H}_{\rm sia}^D$ implies that the $\zeta_i$ axis is a local hard magnetization axis at every site, and consequently the crystal $c$ axis becomes a hard magnetization axis since the $\zeta_i$ axis at each site directs nearly along the $c$. On the other hand, the latter part $\mathcal{H}_{\rm sia}^E$ implies that the $\xi_i$ and $\eta_i$ axes become a local hard magnetization axis alternately in the $ab$ plane. This is due to the staggered ordering of occupied $e_g$ orbitals, i.e. the 3$x^2$-$r^2$/3$y^2$-$r^2$ type orbital ordering.

We derive directional vectors of the $\xi_i$, $\eta_i$ and $\zeta_i$ axes with respect to the $a$, $b$ and $c$ axes using coordination parameters of oxygens as
%%%%%%%%%%%%%%%%%%%%%%%%%%%%%%%%%%%
\begin{eqnarray}
\label{eq:tilax1}
\bm {\xi}_i&=&\left[
\begin{array}{c}
a[0.25+(-1)^{i_x+i_y}(0.75-x_{{\rm O}_2})] \\
b[0.25-(-1)^{i_x+i_y}(y_{{\rm O}_2}-0.25)] \\
c(-1)^{i_x+i_y+i_z}z_{{\rm O}_2} \\
\end{array}
\right], \\
\label{eq:tilax2}
\bm {\eta}_i&=&\left[
\begin{array}{c}
a[-0.25+(-1)^{i_x+i_y}(0.75-x_{{\rm O}_2})] \\
b[0.25+(-1)^{i_x+i_y}(y_{{\rm O}_2}-0.25)] \\
-c(-1)^{i_x+i_y+i_z}z_{{\rm O}_2} \\
\end{array}
\right], \\
\label{eq:tilax3}
\bm {\zeta}_i&=&\left[
\begin{array}{c}
-a(-1)^{i_x+i_y+i_z}x_{{\rm O}_1} \\
b(-1)^{i_z}(0.5-y_{{\rm O}_1}) \\
0.25c \\
\end{array}
\right].
\end{eqnarray}
%%%%%%%%%%%%%%%%%%%%%%%%%%%%%%%%%%%
Here $x_{{\rm O}_2}$, $y_{{\rm O}_2}$ and $z_{{\rm O}_2}$ are the coordination parameters of the in-plane oxygens, $x_{{\rm O}_1}$, $y_{{\rm O}_1}$ and $z_{{\rm O}_1}$ are those of the out-of-plane oxygens, and $a$, $b$ and $c$ are the lattice parameters. For values of these parameters, we use the experimental data of EuMnO$_3$~\cite{Dabrowski05} (see Table~\ref{tabl:SDEMO}) throughout the calculations for simplicity. We have confirmed that the results are not significantly changed even if we use the values for other $R$MnO$_3$ compounds especially with $R$=Gd, Tb, Dy, Y and Ho. We note that the axis vectors $\bm {\xi}_i$, $\bm {\eta}_i$ and $\bm {\zeta}_i$ in Eqs.~(\ref{eq:tilax1})-(\ref{eq:tilax3}) are not normalized.

Using these vectors, we can rewrite $\mathcal{H}_{\rm sia}^D$ and $\mathcal{H}_{\rm sia}^E$ as
%%%%%%%%%%%%%%%%%%%%%%%%%%%%%%%%%%%%%%%%%%%%%%%%%%
\begin{eqnarray}
\mathcal{H}_{\rm sia}^D&=&
D\sum_{i} (\bm S_i \cdot \bm {\zeta}_i/|\bm {\zeta}_i|)^2, \\
\mathcal{H}_{\rm sia}^E&=&
E\sum_{i} (-1)^{i_x+i_y} \\ \nonumber
&\times&[(\bm S_i \cdot \bm {\xi}_i/|\bm {\xi}_i|)^2
-(\bm S_i \cdot \bm {\eta}_i/|\bm {\eta}_i|)^2].
\end{eqnarray}
%%%%%%%%%%%%%%%%%%%%%%%%%%%%%%%%%%%%%%%%%%%%%%%%%%
In Table.~\ref{tabl:PRMVL}, the values of $D$ and $E$ are listed, which are calculated using formulae given in Ref.~\cite{Matsumoto70}. Noticeably, the values are all approximately equal in the compounds located near/in the multiferroic phases in the phase diagram although the calculated values contain some ambiguities due to uncertainty in the optical absorption data~\cite{Matsumoto70,Gerristen63}. We take $D$=0.25 meV and $E$=0.30 meV in the following calculations. For details of the parameter calculations here, see Appendix C.

\subsection{Dzyaloshinsky-Moriya interaction}
\label{SSec:DMint}
%%%%%%%%%%%%%%%%%%%%%%%%%%%%%%%%%%%%%%%%%%%%%%%%%%%%%%%%%%%%
\begin{figure}[tdp]
\includegraphics[scale=1.0]{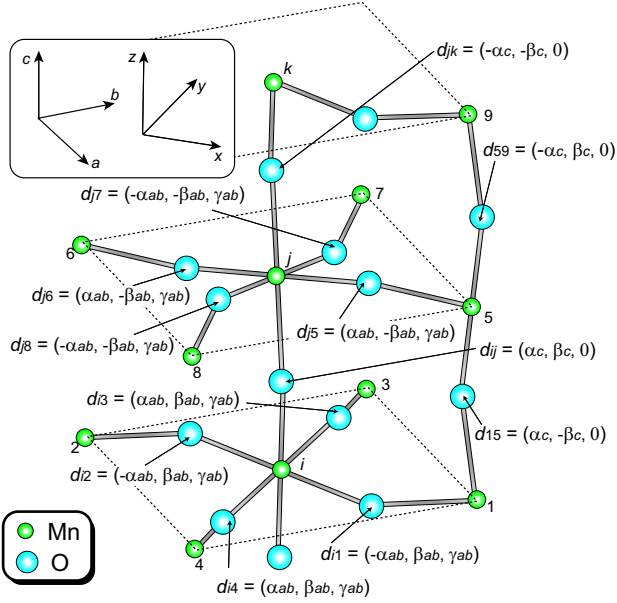}
\caption{(Color online) Dzyaloshinsky-Moriya vectors associated with different Mn-O-Mn bonds, which are expressed by five parameters, $\alpha_{ab}$, $\beta_{ab}$, $\gamma_{ab}$, $\alpha_c$ and $\beta_c$. The vectors on the in-plane Mn-O-Mn bonds direct nearly perpendicular to the plane consisting of the corresponding Mn, O and Mn ions.}
\label{Fig05}
\end{figure}
%%%%%%%%%%%%%%%%%%%%%%%%%%%%%%%%%%%%%%%%%%%%%%%%%%%%%%%%%%%%%
The third term $\mathcal{H}_{\rm DM}$ denotes the Dzyaloshinsky-Moriya (DM) interactions~\cite{Dzyaloshinsky58,Moriya60a,Moriya60b}. The DM vector $\bm d^{\alpha}_{ij}$ is defined on the Mn($i$)-O-Mn($j$) bond along the $\alpha$ axis ($\alpha$=$x$, $y$ and $z$), and the following antisymmetric relation holds; $\bm d^{\alpha}_{ji}=-\bm d^{\alpha}_{ij}$. Given that the Mn($j$) ion is adjacent to the Mn($i$) ion in the positive $\alpha$ direction, the DM vectors are given by (see also Fig.~\ref{Fig05})
%%%%%%%%%%%%%%%%%%%%%%%%%%%%%%%%%%%
\begin{eqnarray}
\bm d_{ij}^x&=&\left[
\begin{array}{c}
(-1)^{i_x+i_y+i_z}\alpha_{ab} \\
-(-1)^{i_x+i_y+i_z}\beta_{ab} \\
(-1)^{i_x+i_y}\gamma_{ab} \\
\end{array}
\right], \\
\bm d_{ij}^y&=&\left[
\begin{array}{c}
-(-1)^{i_x+i_y+i_z}\alpha_{ab} \\
-(-1)^{i_x+i_y+i_z}\beta_{ab} \\
(-1)^{i_x+i_y}\gamma_{ab} \\
\end{array}
\right], \\
\bm d_{ij}^z&=&\left[
\begin{array}{c}
-(-1)^{i_z}\alpha_c \\
-(-1)^{i_x+i_y+i_z}\beta_c \\
0 \\
\end{array}
\right],
\label{eq:DMVECS}
\end{eqnarray}
%%%%%%%%%%%%%%%%%%%%%%%%%%%%%%%%%%%
because of the crystal symmetry.

The five DM parameters, $\alpha_{ab}$, $\beta_{ab}$, $\gamma_{ab}$, $\alpha_c$, and $\beta_c$, in LaMnO$_3$ was evaluated in the first-principles calculation~\cite{Solovyev96}, which shows that the DM vector on the in-plane Mn-O-Mn bond is nearly perpendicular to the plane consisting of the corresponding Mn, O, and Mn ions, while the DM vector on the out-of-plane bond is not. This can be understood as follows. The spin exchange on the in-plane Mn-Mn bond is governed by the exchange path via the inbetween one oxygen, i.e. the Mn-O-Mn path. Thus the direction of the DM vector on the in-plane bond is dominantly determined by local symmetry of the Mn-O-Mn bond. On the other hand, the spin exchange on the out-of-plane Mn-Mn bond is not necessarily governed by the Mn-O-Mn path along the $c$ axis, and contributions from indirect paths containing more than one oxygens are not negligible since the occupied $e_g$ orbitals are directed not along the $c$ axis but in the $ab$ plane. As a result, the DM vector on the out-of-plane bond does not reflect the local symmetry of the Mn-O-Mn bond. 

In addition, according to the electron-spin resonance (ESR) measurements for LaMnO$_3$~\cite{Huber99,Tovar99,Deisenhofer02}, the DM vector on the out-of-plane bond is approximately four times larger in magnitude than the vector on the in-plane bond. We expect that the above findings also hold in other $R$MnO$_3$ compounds although neither theoretical nor experimental studies on the DM vectors presently exists except for $R$=La. In our calculation, we take $\alpha_{ab}$=0.10 meV, $\beta_{ab}$=0.10 meV, $\gamma_{ab}$=0.14 meV, $\alpha_c$=0.30 meV and $\beta_c$=0.30 meV. With this set of parameters, the above two features are reproduced: the DM vector on the in-plane bond is nearly perpendicular to the corresponding Mn-O-Mn plane, and the vector on the out-of-plane bond is approximately three times larger than the vector on the in-plane bond.

We also mention that the $a$ components of DM vectors on the out-of-plane bonds have the same signs within each plane, but the signs alternate along the $c$ axis. This situation gives rise to a weak ferromagnetism with moments along the $c$ axis due to the spin canting in the AFM(A) phase in agreement with the experimental observation~\cite{Yamasaki07b,Skumryev99}.

\subsection{Cubic anisotropy}
\label{SSec:CubicAniso}
The last term $\mathcal{H}_{\rm cub}$ represents the cubic anisotropy, which comes from nearly cubic symmetry of the perovskite lattice. Here, $S_x$, $S_y$ and $S_z$ are written as $S_x=1/\sqrt{2}(S_b-S_a)$, $S_y=1/\sqrt{2}(S_b+S_a)$, and $S_z=S_c$. The coupling constant $a$ in the Mn$^{3+}$ ion surrounded by the octahedrally coordinated oxygens was evaluated to be 0.0162 meV in the ESR measurement~\cite{Gerristen63}. We neglect a slight contribution from the orthorhombic lattice distortion to the spin anisotropy since it is expected to be very small.

\subsection{Method}
\label{SSec:Method}
%%%%%%%%%%%%%%%%%%%%%%%%%%%%%%%%%%%%%%%%%%%%%%%%%%%%%%%%%%%%%%%%%%
\begin{table}
\caption{Model parameters used in the calculations for each term of the Hamiltonian. The energy unit is meV.}
\begin{tabular}{c|cccc}
\hline
$\mathcal{H}_{\rm ex}$  & $J_{ab}$=0.80, & $J_c$=1.25 & & \\
$\mathcal{H}_{\rm sia}$ & $D$=0.25,      & $E$=0.30   & & \\
$\mathcal{H}_{\rm DM}$  &
$\alpha_{ab}$=0.10, & $\beta_{ab}$=0.10, & $\gamma_{ab}$=0.14 \\
                        &
$\alpha_c$=0.30, & $\beta_c$=0.30, & $\gamma_c$=0.0 \\
$\mathcal{H}_{\rm cub}$ & $a$=0.0162    &            & & \\
\hline
\end{tabular}
\label{tabl:MDLPRMS}
\end{table}
%%%%%%%%%%%%%%%%%%%%%%%%%%%%%%%%%%%%%%%%%%%%%%%%%%%%%%%%%%%%%%%%%%
We calculate thermodynamic properties of the model (\ref{eq:model1}) by using the Monte-Carlo method. The model parameters used in the calculations are summarized in Table~\ref{tabl:MDLPRMS}.
To avoid a critical slowing down in the frustrated systems, 
% To overcome the slow MC dynamics at low temperatures, 
we employ the replica exchange Monte-Carlo method~\cite{Hukushima96}. We take an exchange sampling after every 400 standard Monte-Carlo steps. Typically, we perform 600 exchanges after sufficient thermalization Monte-Carlo steps. In the following, we mainly show the results obtained for systems with 48$\times$48$\times$6 sites under the periodic boundary condition. By performing the calculations also for systems with 36$\times$36$\times$6 sites at some parameter values, and by adopting the open boundary condition also, we confirm that the finite-size effect is small enough and never affects our conclusion.

On the other hand, we study ground-state properties by numerically solving the Landau-Lifshitz-Gilbert equation. We derive effective local magnetic fields $\bm H^{\rm eff}_i$ acting on the $i$-th Mn spin $\bm S_i$ from the Hamiltonian $\mathcal{H}$ as
%%%%%%%%%%%%%%%%%%%%%%%%%%%%%%%%%%%
\begin{equation}
\bm H^{\rm eff}_i = - \partial \mathcal{H} / \partial \bm S_i.
\label{eq:EFFMF}
\end{equation}
%%%%%%%%%%%%%%%%%%%%%%%%%%%%%%%%%%%
Then, we construct a Landau-Lifshitz-Gilbert equation with thus obtained local fields;
%%%%%%%%%%%%%%%%%%%%%%%%%%%%%%%%%%%
\begin{equation}
\frac{\partial \bm S_i}{\partial t}=-\bm S_i \times \bm H^{\rm eff}_i
+ \frac{\alpha_{\rm G}}{S} \bm S_i \times \frac{\partial \bm S_i}{\partial t},
\label{eq:LLGEQ}
\end{equation} 
%%%%%%%%%%%%%%%%%%%%%%%%%%%%%%%%%%%
where $\alpha_{\rm G}$ is the dimensionless Gilbert-damping coefficient introduced phenomenologically. For the value of $\alpha_{\rm G}$, we take a rather small value of $\alpha_{\rm G}=0.01$ to achieve a slow relaxation towards a stable spin structure with a minimum energy. We solve this equation using the Runge-Kutta method. For the convergence, we use thermally relaxed spin configurations obtained in the Monte-Carlo simulations at the lowest temperature ($k_{\rm B}T\sim$~0.5 meV) as initial states.

\section{Results}
\label{Sec:Results}
\subsection{Overview}
\label{SSec:Overview}
When we study the experimental phase diagrams introduced in Sec.~\ref{Sec:Intro}, we assume that main roles of the GdFeO$_3$-type distortion on the magnetic and electric properties in $R$MnO$_3$ are induction and enhancement of the second-neighbor antiferromagnetic exchanges $J_2$. Under this assumption, we investigate the $T$-$r_R$ diagrams for $R$MnO$_3$ ($r_R$ is the ionic $R$-site radius) and the $T$-$x$ diagrams for solid solutions by varying the value of $J_2$. In Sec~\ref{SSec:PhsDgm}, we first display a theoretically obtained $T$-$J_2$ phase diagram, which shows good agreement with the experimentally obtained $T$-$x$ phase diagram of Eu$_{1-x}$Y$_x$MnO$_3$ in Ref.~\cite{Yamasaki07b}. Then we show calculated results for the momentum dependence of spin and spin-helicity correlation functions, and the temperature dependence of specific heat and spin-helicity vector. We discuss how we identify phase transitions and magnetic structures in the phase diagram with the aid of these results. In Sec.~\ref{SSec:PFlop}, we discuss a mechanism of the orthorhombic-distortion-induced spin-cycloidal-plane (electric-polarization) flop. We also demonstrate that the regime of $bc$-cycloidal spin ($P$$\parallel$$c$) phase increases as the DM parameter $\alpha_c$ increases, and the experimental phase diagram of the Gd$_{1-x}$Tb$_x$MnO$_3$ system in Ref.~\cite{Goto05} is reproduced well for a rather large value of $\alpha_c$=~0.38 meV. In Sec.~\ref{SSec:Ellipticity}, we study the ground-state spin structures in the $ab$- and $bc$-cycloidal spin phases, and discuss the experimentally claimed elliptical modulation of the cycloidal spin structure. In Sec.~\ref{SSec:SIAAFMA}, by examining effects of the single-ion anisotropies, we address the issue why the Mn spins are aligned along the $b$ axis in the AFM(A) and sinusoidal collinear phases. In Sec.~\ref{SSec:SinColl}, we discuss the sinusoidal collinear spin phase in the intermediate temperature region by examining the role of the single-ion anisotropies on its emergence.

\subsection{Phase diagram}
\label{SSec:PhsDgm}
%%%%%%%%%%%%%%%%%%%%%%%%%%%%%%%%%%%%%%%%%%%%%%%%%%%%%%%%%%%%
\begin{figure}[tdp]
\includegraphics[scale=1.0]{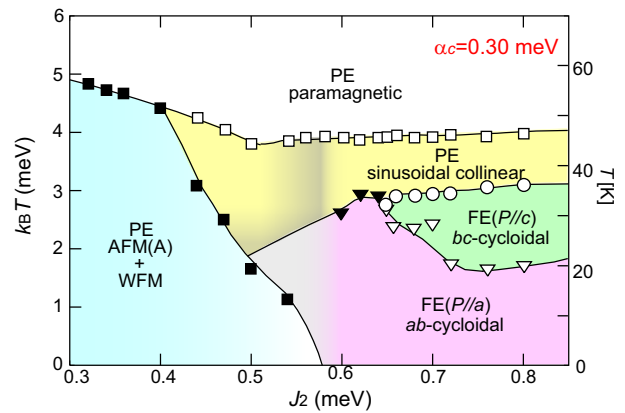}
\caption{(Color online) Theoretically obtained magnetoelectric phase diagram in plane of temperature and $J_2$. Here, AFM(A)+WFM denotes the A-type antiferromagnetic phase with weak ferromagnetism due to the spin canting, and PE and FE denote paraelectric and ferroelectric phases expected in the spin-current model~\cite{Katsura05,Sergienko06a,Mostovoy06}, respectively.
%%, which reproduce the experimental phase diagrams for 
%%Eu$_{1-x}$Y$_x$MnO$_3$ and Gd$1-x$Tb$_x$MnO$_3$, respectively.
}
\label{Fig06}
\end{figure}
%%%%%%%%%%%%%%%%%%%%%%%%%%%%%%%%%%%%%%%%%%%%%%%%%%%%%%%%%%%%%
In Fig.~\ref{Fig06}, we display a theoretically obtained $T$-$J_2$ phase diagram, which reproduces the experimental $T$-$x$ phase diagram of Eu$_{1-x}$Y$_x$MnO$_3$ system [Fig.~\ref{Fig02}(b)]. This theoretical diagram shows that at the lowest temperature, the canted AFM(A) state emerges for $J_2\le$~0.59, while for $J_2\ge$~0.59, the cycloidal spin states emerge. For 0.59$\le J_2 \le$~0.64, the system undergoes two thermal phase transitions, and the paramagnetic, sinusoidal collinear, and $ab$-cycloidal phases successively emerge with lowering temperature. On the other hand, for $J_2\ge$~0.64, the system undergoes three thermal phase transitions, and four magnetic phases, i.e., the paramagnetic, sinusoidal collinear, $ab$-cycloidal and $bc$-cycloidal phases, successively emerge as temperature decreases. The third transition is nothing but a spin-cycloidal-plane flop from the $bc$ plane to the $ab$ plane. The regime of $bc$-cycloidal phase increases with increasing $J_2$. According to the spin-current model, the ferroelectricity with $P_a$ is expected to show up in the $ab$-cycloidal spin phase and that with $P_c$ in the $bc$-cycloidal spin phase. Note that in the gray region in Fig.~\ref{Fig06}, we obtain several complicated magnetic structures, whose spin correlation functions show peaks at several k-points. They are considered to be artifacts of the finite-size calculation for 48$\times$48$\times$6 sites. In the vicinity of the AFM(A) phase, pitches of the spiral and sinusoidal spin states become so long and the magnetic unit cells become so large that the calculation tends to be seriously affected by the finite-size effects.

We identify each of the phases and each of the transitions in the above phase diagram by calculating several physical quantities. The magnetic structures are assigned from calculated momentum dependence of the spin correlation function $\hat{S}_{\alpha}(\bm k,T)$ and that of the spin-helicity correlation function $\hat{H}_{\alpha}^b(\bm k,T)$. On the other hand, the transition points are determined from calculated temperature dependence of the specific heat $C_s(T)$ and that of the spin-helicity vector $\bm h_s^b(T)$.

%%%%%%%%%%%%%%%%%%%%%%%%%%%%%%%%%%%%%%%%%%%%%%%%%%%%%%%%%%%%%
\begin{figure*}[tdp]
\includegraphics[scale=1.0]{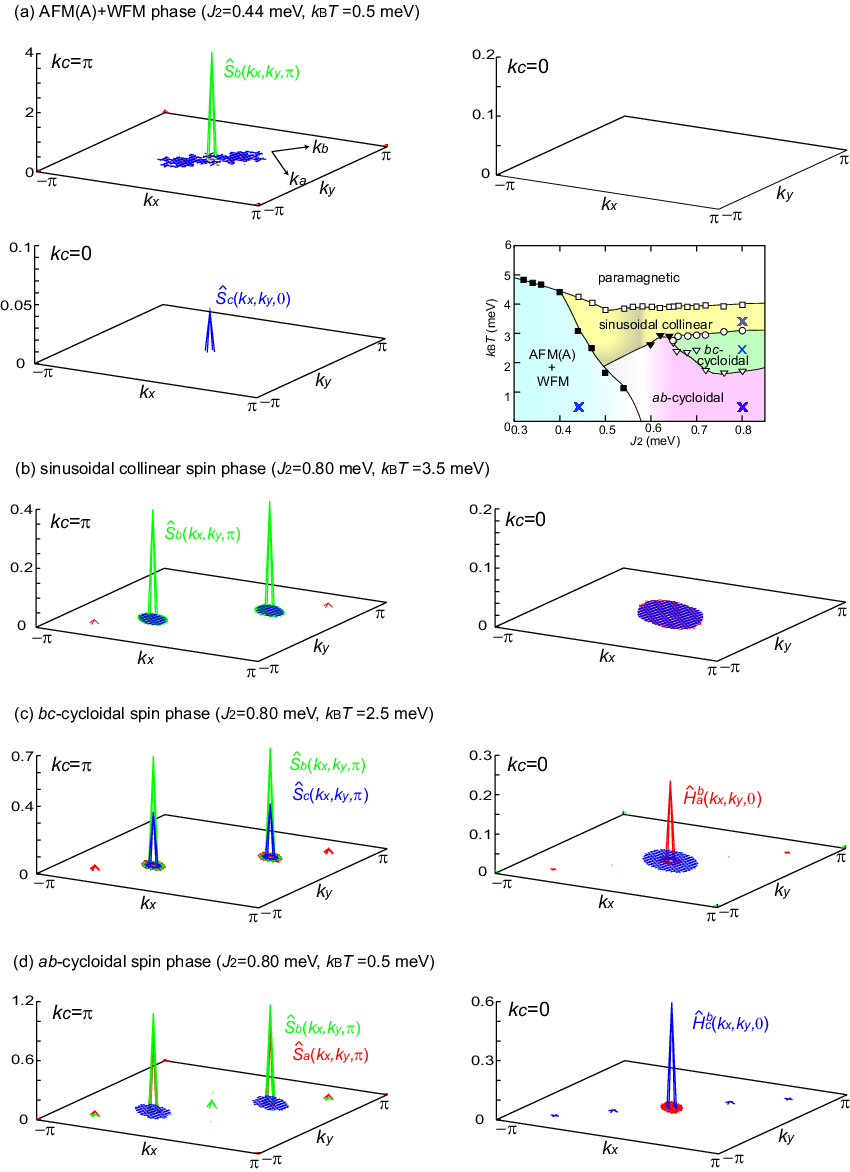}
\caption{(Color online) Calculated spin correlation functions $\hat{S}_{\alpha}(\bm k)$ for $k_c$=$\pi$ (left panels) and spin-helicity correlation functions $\hat{H}_{\alpha}^b(\bm k)$ for $k_c$=0 (right panels) in the $k_x$-$k_y$ plane ($\alpha$=$a$, $b$ and $c$) for different magnetic states ----- at the points indicated by cross symbols in the inset; (a) the canted AFM(A) state [AFM(A)+WFM state] at $J_2$=0.44 meV and $k_{\rm B}T$=0.5 meV, (b) the sinusoidal collinear spin state at $J_2$=0.80 meV and $k_{\rm B}T$=3.5 meV, (c) the $bc$-cycloidal spin state at $J_2$=0.80 meV and $k_{\rm B}T$=2.5 meV, and (d) the $ab$-cycloidal spin state at $J_2$=0.80 meV and $k_{\rm B}T$=0.5 meV. For the AFM(A)+WFM state, spin correlation function $\hat{S}_{\alpha}(\bm k)$ for $k_c$=0 is also displayed, which indicates small ferromagnetic moments along the $c$ axis.}
\label{Fig07}
\end{figure*}
%%%%%%%%%%%%%%%%%%%%%%%%%%%%%%%%%%%%%%%%%%%%%%%%%%%%%%%%%%%%%%
The spin and spin-helicity correlation functions $\hat{S}_{\alpha}(\bm k,T)$ and $\hat{H}_{\alpha}^b(\bm k,T)$ for $\alpha$=$a$, $b$, $c$ are calculated by
%%%%%%%%%%%%%%%%%%%%%%%%%%%%%%%%%%%%%%%%%%%%%%%%%%
\begin{equation}
\hat{S}_{\alpha}(\bm k,T)=\frac{1}{N^2} 
\sum_{i,j} \langle S_{\alpha i}S_{\alpha j}\rangle
e^{i\bm k \cdot (\bm r_i-\bm r_j)},
\label{eq:scrrf}
\end{equation}
\begin{equation}
\hat{H}_{\alpha}^b(\bm k,T)=\frac{1}{N^2} 
\sum_{i,j} \langle h_{\alpha i}^bh_{\alpha j}^b\rangle
e^{i\bm k \cdot (\bm r_i-\bm r_j)},
\label{eq:hcrrf}
\end{equation}
%%%%%%%%%%%%%%%%%%%%%%%%%%%%%%%%%%%%%%%%%%%%%%%%%%
where the bracket denotes the thermal average.
Here, $h_{\alpha i}^b$ is the $\alpha$ component of local spin-helicity vector $\bm h_i^b=(h_{a i}^b, h_{b i}^b, h_{c i}^b)$, which is defined as
%%%%%%%%%%%%%%%%%%%%%%%%%%%%%%%%%%%%%%%%%%%%%%%%%%
\begin{equation}
\bm h_i^b=(\bm S_i \times \bm S_{i+b})/S^2.
\label{eq:hlcty2}
\end{equation}
%%%%%%%%%%%%%%%%%%%%%%%%%%%%%%%%%%%%%%%%%%%%%%%%%%
In the following, we write these correlation functions simply as $\hat{S}_{\alpha}(\bm k)$ and $\hat{H}_{\alpha}^b(\bm k)$ by omitting $T$.

In a cycloidal spin state, all of the local spin-helicity vectors point in the same direction perpendicular to the basal cycloidal plane. In this sense, the cycloidal spin state can be regarded as a $\it ferrohelicity$ state. Furthermore, the $ab$-cycloidal spin state has a ferrohelicity component along the $c$ axis, while the $bc$-cycloidal spin state has that along the $a$ axis. Thus the spin-helicity $c$-component correlation $\hat{H}_c^b(\bm k)$ has a peak at $\bm k$=(0, 0, 0) in the $ab$-cycloidal spin phase, while in the $bc$-cycloidal spin phase, its $a$-component correlation $\hat{H}_a^b(\bm k)$ has a peak also at $\bm k$=(0, 0, 0). In the following, we overview features of $\hat{S}_{\alpha}(\bm k)$ and $\hat{H}_{\alpha}^b(\bm k)$ in each magnetic phase by taking typical points in the phase diagram as examples, which are indicated by cross symbols in the inset of Fig.~\ref{Fig07}(a). 

In Fig.~\ref{Fig07}(a) we display the calculated $\hat{S}_{\alpha}(k_x, k_y, \pi)$ (upper left panel) and $\hat{H}_{\alpha}^b(k_x, k_y, 0)$ (right panel) in the AFM(A)+WFM phase. In this phase, the Mn spins are aligned nearly parallel to the $b$ axis, and couple ferromagnetically in the $ab$ plane with antiferromagnetic stacking along the $c$ axis. As a result, the spin $b$-component correlation $\hat{S}_b(\bm k)$ has a sharp peak at $\bm k=(0, 0, \pi)$, while $\hat{S}_a(\bm k)$ and $\hat{S}_c(\bm k)$ do not. On the other hand, the spin-helicity correlation $\hat{H}_{\alpha}(\bm k)$ has no structure for any $\alpha$. In addition, weak ferromagnetic moments arise due to the spin canting. As shown in the lower panel of Fig.~\ref{Fig07}(a), the spin $c$-component correlation $\hat{S}_c(\bm k)$ exhibits a tiny but sharp peak at $\bm k=0$. Moreover, this spin canting in fact generates finite local spin helicities $\bm h_i^c$ between spin pairs neighboring along the $c$ axis, defined as 
%%%%%%%%%%%%%%%%%%%%%%%%%%%%%%%%%%%%%%%%%%%%%%%%%%
\begin{equation}
\bm h_i^c=(\bm S_i \times \bm S_{i+c})/S^2.
\label{eq:hlcty3}
\end{equation}
%%%%%%%%%%%%%%%%%%%%%%%%%%%%%%%%%%%%%%%%%%%%%%%%%%
These vectors direct in the positive $a$ direction or in the negative $a$ direction uniformly in the plane, and these two kinds of planes are stacked alternately. Thus if we introduce the correlation function of $\bm h_i^c$ as
%%%%%%%%%%%%%%%%%%%%%%%%%%%%%%%%%%%%%%%%%%%%%%%%%%
\begin{equation}
\hat{H}_{\alpha}^c(\bm k,T)=\frac{1}{N^2} 
\sum_{i,j} \langle h_{\alpha i}^ch_{\alpha j}^c\rangle
e^{i\bm k \cdot (\bm r_i-\bm r_j)},
\label{eq:hcrrfC}
\end{equation}
%%%%%%%%%%%%%%%%%%%%%%%%%%%%%%%%%%%%%%%%%%%%%%%%%%
the $a$-component correlation $\hat{H}_a^c(\bm k,T)$ has a peak at $\bm k$=(0, 0, $\pi$) ----- not shown.

In Fig.~\ref{Fig07}(b), we display the calculated $\hat{S}_{\alpha}(k_x, k_y, \pi)$ and $\hat{H}_{\alpha}^b(k_x, k_y, 0)$ in the $ab$-cycloidal phase. In this phase, the Mn spins rotate in the $ab$ plane, while they couple antiferromagnetically along the $c$ axis. As a result, the spin correlations $\hat{S}_a(\bm k)$ and $\hat{S}_b(\bm k)$ exhibit sharp peaks at $(k_a, k_b, k_c)=(0, \pm q_m, \pi)$ with a finite $q_m$. The rotating spins in the $ab$ plane give rise to a ferro-arrangement of the spin helicities $\bm h_i^b$ along the $c$ axis. This results in a peak of spin-helicity correlation $\hat{H}_c^b(\bm k)$ at $\bm k$=0.

The correlation functions in the $bc$-cycloidal phase in Fig.~\ref{Fig07}(c) exhibit that the spin correlations $\hat{S}_b(\bm k)$ and $\hat{S}_c(\bm k)$ show sharp peaks at $(k_a, k_b, k_c)=(0, \pm q_m, \pi)$, and the spin-helicity correlation $\hat{H}_a^b(\bm k)$ has a sharp peak at $k=0$. This is because the Mn spins rotating in the $bc$ plane give rise to a ferro-arrangement of the spin helicities $\bm h_i^b$ along the $a$ axis.

Finally, the results for sinusoidal collinear phase are shown in Fig.~\ref{Fig07}(d). In this phase, the collinearly aligned Mn spins ($\parallel$$b$) are sinusoidally modulated with a finite modulation vector along the $b$ axis. This results in sharp peaks of the spin $b$-component correlation $\hat{S}_b(\bm k)$ at $(k_a, k_b, k_c)=(0, \pm q_m, \pi)$ with a finite $q_m$, as well as no remarkable structure in the spin-helicity correlations $\hat{H}_{\alpha}^b(\bm k)$ for any $\alpha$.

%%%%%%%%%%%%%%%%%%%%%%%%%%%%%%%%%%%%%%%%%%%%%%%%%%%%%%%%%%%%%
\begin{figure}[tdp]
\includegraphics[scale=1.0]{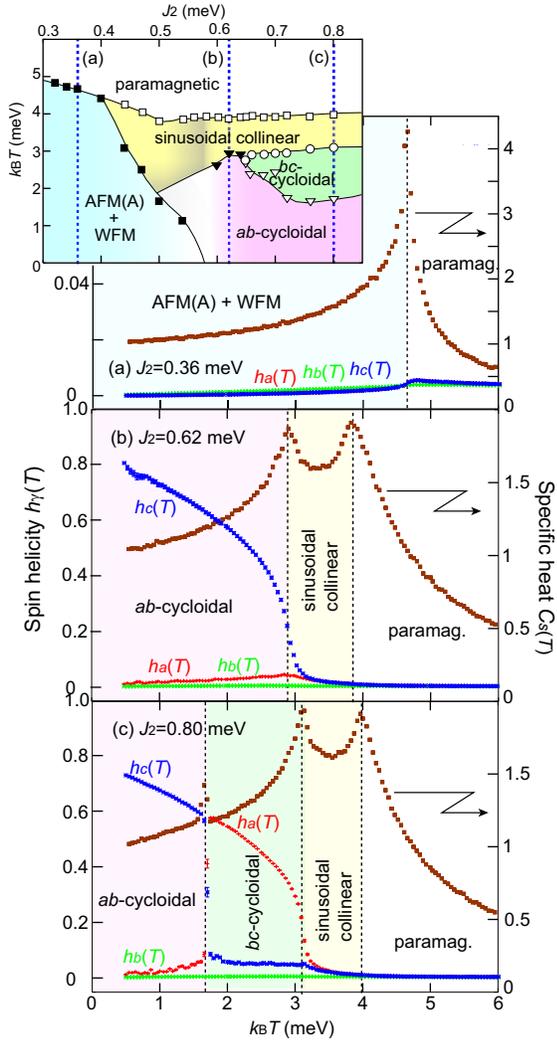}
\caption{(Color online) Calculated temperature profiles of specific heat $C_s(T)$ and spin-helicity vector $\bm h_s^b(T)$ for various values of $J_2$ (along the vertical dashed lines in inset); (a) $J_2$=0.36 meV, (b) $J_2$=0.62 meV, and (c) $J_2$=0.80 meV. Here, $h_\gamma(T)$ ($\gamma$=$a$, $b$, $c$) denotes the $\gamma$ component of $\bm h_s^b(T)$. In the $ab$-cycloidal [$bc$-cycloidal] phase, the value of $h_c(T)$ [$h_a(T)$] is much larger than other two components.}
\label{Fig08}
\end{figure}
%%%%%%%%%%%%%%%%%%%%%%%%%%%%%%%%%%%%%%%%%%%%%%%%%%%%%%%%%%%%%%
In Fig.~\ref{Fig08}, we show the temperature dependence of specific heat $C_s(T)$ and spin-helicity vector $\bm h_s^b(T)$ for various values of $J_2$, which are calculated by,
%%%%%%%%%%%%%%%%%%%%%%%%%%%%%%%%%%%%%%%%%%%%%%%%%%
\begin{eqnarray}
\label{eq:spheat}
C_s(T)&=&\frac{1}{N} \partial \langle \mathcal{H}\rangle 
/ \partial (k_{\rm B}T),\\
\label{eq:hlcty1}
\bm h_s^b(T)&=&
\frac{1}{N} \langle |\sum_{i} \bm S_i \times \bm S_{i+b} |\rangle/S^2.
\end{eqnarray}
%%%%%%%%%%%%%%%%%%%%%%%%%%%%%%%%%%%%%%%%%%%%%%%%%%
Here, $h_\gamma(T)$ ($\gamma$=$a$, $b$ and $c$) denotes the $\gamma$ component of spin-helicity vector $\bm h_s^b(T)=[h_a(T), h_b(T), h_c(T)]$. In the $ab$-cycloidal [$bc$-cycloidal] phase, its $c$ [$a$] component $h_c(T)$ [$h_a(T)$] has a large value, while other two components are strongly suppressed. On the other hand, in the paramagnetic, canted AFM(A) and sinusoidal collinear phases, all of these three components should be nearly equal to zero.

In Fig.~\ref{Fig08}(a), we can see a single phase transition in $C_s(T)$ for a small value of $J_2$=0.36 meV, which corresponds to a transition from paramagnetic into canted AFM(A) phases. The spin-helicity vector $\bm h_s^b(T)$ is nearly equal to zero constantly through this transition.

For a larger value of $J_2$=0.62 meV, $C_s(T)$ shows two peaks as shown in Fig.~\ref{Fig08}(b), indicating that successive two transitions occur with lowering temperature. The spin-helicity vector $\bm h_s^b(T)$ is almost unchanged, and is approximately zero through the first transition, at which the system enters into the sinusoidal collinear phase from paramagnetic phase. Contrastingly, at the subsequent transition, its $c$ component $h_c(T)$ starts increasing, indicating a transition into the $ab$-cycloidal phase.

For a further increased value of $J_2$=0.80 meV, the system exhibits successive three phase transitions as temperature decreases ----- see Fig.~\ref{Fig08}(c). The first one is a transition from paramagnetic to sinusoidal collinear phases through which all of the three components of $\bm h_s^b(T)$ are again approximately zero constantly. At the second transition, its $a$ component, $h_a(T)$, increases, while other two components are remain to be small, indicating a transition into the $bc$-cycloidal phase. With further lowering temperature, the $a$ component $h_a(T)$ suddenly drops, while the $c$ component $h_c(T)$ steeply increases at the third transition point, indicating a spin-cycloidal-plane flop from the $bc$ plane to the $ab$ plane.

\subsection{Orthorhombic-distortion-induced polarization flop}
\label{SSec:PFlop}

In $R$MnO$_3$, the $c$ axis is always a hard axis for magnetization due to the single-ion anisotropies $\mathcal{H}_{\rm sia}^D$. Therefore, the $bc$-cycloidal spin state seems to be higher in energy than the $ab$-cycloidal spin state at first sight. In this sense, the emergence of $bc$-cycloidal spin order is puzzling. Moreover, the $bc$-cycloidal spin phase emerges next to the $ab$-cycloidal spin phase with increasing GdFeO$_3$-type distortion. The mechanism of this lattice-distortion-induced cycloidal-plane flop has not been clarified yet.

%%%%%%%%%%%%%%%%%%%%%%%%%%%%%%%%%%%%%%%%%%%%%%%%%%%%%%%%%%%%
\begin{figure}[tdp]
\includegraphics[scale=1.0]{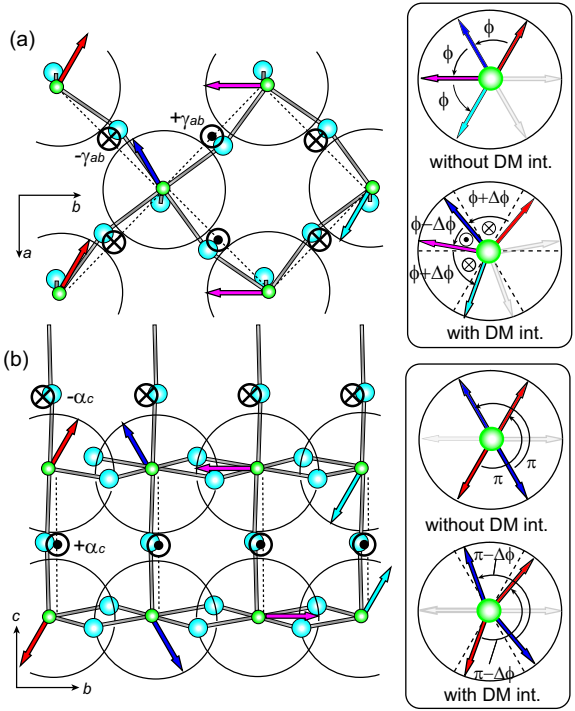}
\caption{(Color online) (a) [(b)] Left panel: Spin structure in the $ab$-cycloidal [$bc$-cycloidal] state and arrangement of the $c$ [$a$] components of DM vectors on the in-plane [out-of-plane] Mn-O-Mn bonds. The symbols $\odot$ and $\otimes$ express their signs, i.e. $\odot$ for the positive sign and $\otimes$ for the negative sign. Right panel: (upper panel) Spin directions in the $ab$-cycloidal [$bc$-cycloidal] state with uniform rotation angles in the absence of DM interactions, and (lower panel) spin directions with modulated rotation angles in the presence of DM interactions.}
\label{Fig09}
\end{figure}
%%%%%%%%%%%%%%%%%%%%%%%%%%%%%%%%%%%%%%%%%%%%%%%%%%%%%%%%%%%%%
If the Hamiltonian contains the superexchange term only, spins in the cycloidal state rotate uniformly. In the ground state, the angle $\phi$ between adjacent two spins along the $x$ or $y$ axis satisfies the relation, $\cos\phi=J_{ab}/(2J_2)$. On the other hand, if we introduce the DM interactions, the rotation angles should be modulated, and are no longer the same.

In the $ab$-cycloidal spin state, the rotating spins couple dominantly to the $c$ components of DM vectors on the in-plane Mn-O-Mn bonds. Their magnitudes are all equal to $\gamma_{ab}$, and their signs (i.e. $+\gamma_{ab}$ and $-\gamma_{ab}$) alternate along the $x$ and $y$ bonds ----- see the left panel of Fig.~\ref{Fig09}(a). Without DM interaction, the spins rotate with the same rotation angles of $\phi_{ab}$. On the other hand, in the presence of DM interactions, the rotation angles become to be alternately modulated into $\phi_{ab}+\Delta \phi_{ab}$ and $\phi_{ab}-\Delta \phi_{ab}$ with $\Delta \phi_{ab}>0$ to get an energy gain from the DM interactions ----- see the right panel of Fig.~\ref{Fig09}(a). We can derive the energy gain due to this angle modulation as
%%%%%%%%%%%%%%%%%%%%%%%%%%%%%%%%%%%%%%%%%%%%%%%%%%
\begin{eqnarray}
\Delta E_{\rm DM}^{ab}/N
&=&-\gamma_{ab}S^2|\sin(\phi_{ab}-\Delta\phi_{ab})-\sin\phi_{ab}| \nonumber \\
&=&-\gamma_{ab}S^2 |\cos\phi_{ab}| \Delta\phi_{ab}.
\end{eqnarray}
%%%%%%%%%%%%%%%%%%%%%%%%%%%%%%%%%%%%%%%%%%%%%%%%%%
This expression implies that the energy gain $|\Delta E_{\rm DM}^{ab}|$ is reduced with increasing $\phi_{ab}$ because the prefactor $|\cos\phi_{ab}|$ becomes maximum (=1) for $\phi_{ab}$~=0 but decreases as $\phi_{ab}$ increases. Thus the $ab$-cycloidal spin state is destabilized with increasing $J_2$ or with increasing GdFeO$_3$-type distortion.

On the other hand, the spins in the $bc$-cycloidal state dominantly couple to the $a$ components of DM vectors on the out-of-plane Mn-O-Mn bonds. Their magnitudes are all equal to $\alpha_c$, and their signs are the same within a plane, but alternate along the $c$ axis ----- see the left panel of Fig.~\ref{Fig09}(b). Without DM interaction, the angles between adjacent two spins along the $c$ axis are uniformly $\phi_c=\pi$ because of the strong antiferromagnetic coupling $J_c$. In the presence of DM interactions, the angles again suffer from staggered modulation into $\pi+\Delta\phi_c$ and $\pi-\Delta\phi_c$ with $\Delta \phi_c>0$ ----- see the right panel of Fig.~\ref{Fig09}(b). Similarly to the $ab$-cycloidal case, we can derive the energy gain due to the angle modulation as
%%%%%%%%%%%%%%%%%%%%%%%%%%%%%%%%%%%%%%%%%%%%%%%%%%
\begin{eqnarray}
\Delta E_{\rm DM}^{bc}/N
&=&-\alpha_c S^2 |\cos \phi_c| \Delta \phi_c \nonumber \\
&=&-\alpha_c S^2 \Delta \phi_c.
\end{eqnarray}
%%%%%%%%%%%%%%%%%%%%%%%%%%%%%%%%%%%%%%%%%%%%%%%%%%
Since the value of $|\cos\phi_c|$ becomes maximum (=1) at $\phi_c$=~$\pi$, the energy gain $\Delta E_{\rm DM}^{bc}$ for the $bc$-cycloidal spin state is always maximum irrespective of the value of $J_2$.

As a result, the energetical advantage of $bc$-cycloidal spin state relative to the $ab$-cycloidal one due to the DM interactions, $|\Delta E_{\rm DM}^{bc}-\Delta E_{\rm DM}^{ab}|$, increases as $J_2$ increases. The $bc$-cycloidal spin state is expected to be stabilized when the above energy difference dominates over the energetical disadvantage due to the hard magnetization $c$ axis.

%%%%%%%%%%%%%%%%%%%%%%%%%%%%%%%%%%%%%%%%%%%%%%%%%%%%%%%%%%%%
\begin{figure}[tdp]
\includegraphics[scale=1.0]{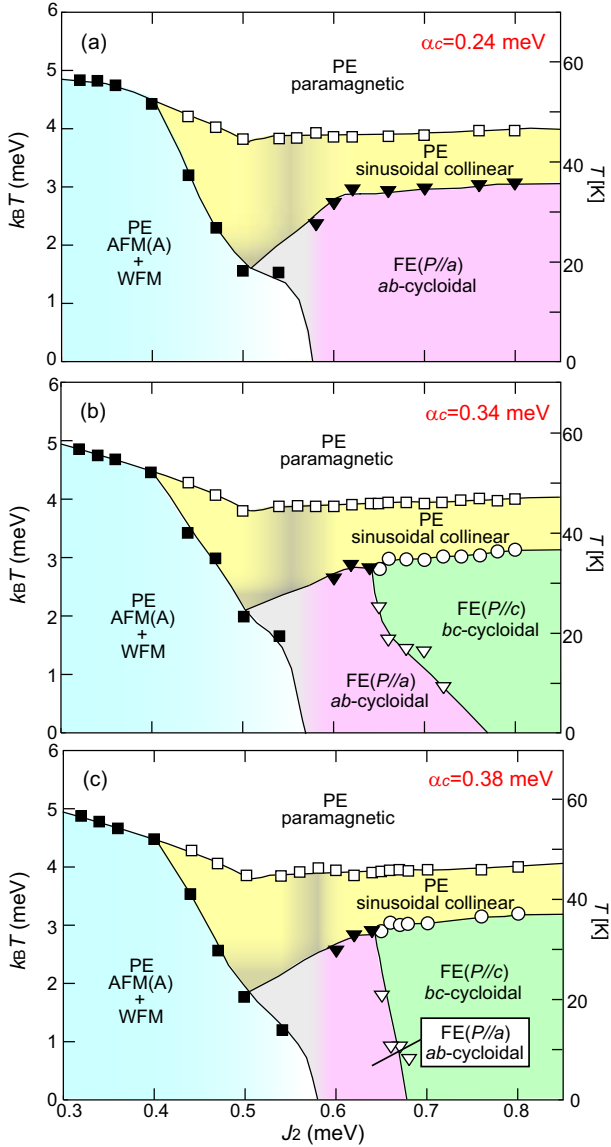}
\caption{(Color online) Theoretically obtained magnetoelectric phase diagrams for (a) $\alpha_c$=0.24 meV, (b) $\alpha_c$=0.34 meV, and (c) $\alpha_c$=0.38 meV. Here $\alpha_c$ expresses magnitude of the $a$ component of DM vector on the out-of-plane Mn-O-Mn bond. PE and FE denote paraelectric and ferroelectric phases expected in the spin-current model~\cite{Katsura05,Sergienko06a,Mostovoy06}, respectively.}
\label{Fig10}
\end{figure}
%%%%%%%%%%%%%%%%%%%%%%%%%%%%%%%%%%%%%%%%%%%%%%%%%%%%%%%%%%%%%
According to the above discussion, we expect that the $a$ component of DM vector on the out-of-plane Mn-O-Mn bond is relevant to the stability of $bc$-cycloidal spin state. To confirm this, we investigate $T$-$J_2$ phase diagrams for various values of $\alpha_c$. Figures~\ref{Fig10}(a)-\ref{Fig10}(c), together with Fig.~\ref{Fig06}, show that the regime of $bc$-cycloidal phase increases as $\alpha_c$ increases. Note that the phase diagram for $\alpha_c$=0.30 meV depicted in Fig.~\ref{Fig06} should be placed between Figs.~\ref{Fig10}(a) and \ref{Fig10}(b), which are for $\alpha_c$=0.24 meV and $\alpha_c$=0.34 meV, respectively.

The experimental phase diagram of Gd$_{1-x}$Tb$_x$MnO$_3$ depicted in Fig.~\ref{Fig02}(b) is reproduced when $\alpha_c$=0.38 meV as in Fig.~\ref{Fig10}(c). This value is slightly larger than $\alpha_c$=0.30 meV, for which the phase diagram of Eu$_{1-x}$Y$_x$MnO$_3$ is reproduced. This difference may be due to different $R$-site-radius dependence of the GdFeO$_3$-type distortion. According to Ref.~\cite{Hemberger07}, the lattice parameters of TbMnO$_3$ are equivalent to those of Eu$_{1-x}$Y$_x$MnO$_3$ with $x\sim0.85$ although TbMnO$_3$ is expected to be located at $x\sim$~0.4 in the $T$-$x$ diagram in terms of the averaged $R$-site radius. This indicates that the lattice of TbMnO$_3$ is more significantly distorted than expected from comparison to the Eu$_{1-x}$Y$_x$MnO$_3$ system ----- we may have to consider not only the average of $R$-site radii but also their variance~\cite{Tomioka04}. The out-of-plane Mn-O-Mn bond angles in Gd$_{1-x}$Tb$_x$MnO$_3$ tend to more significantly deviate from 180$^{\circ}$ than those in Eu$_{1-x}$Y$_x$MnO$_3$ system, resulting in larger DM vectors on the out-of-plane bonds or a larger value of $\alpha_c$. (Note that the DM vector becomes zero when the bond angle is 180$^{\circ}$ because of symmetry.)

We should also note that there is a slight difference between the experimental diagram of Gd$_{1-x}$Tb$_x$MnO$_3$ and theoretical diagram for $\alpha_c$=0.38 meV ----- compare Fig.~\ref{Fig02}(b) and Fig.~\ref{Fig10}(c). In the experimental one, the phase boundary between the $ab$- and $bc$-cycloidal spin phases slightly bends, and in the very narrow region near the phase boundary, the system exhibits a reentrant behavior with successive transitions from $bc$- to $ab$- and again to $bc$-cycloidal spin phases with lowering temperature. In addition, the $ab$-cycloidal spin order is absent in the ground state. These points are not reproduced in our calculation. This discrepancy may be solved by considering effects of the $f$-electron moments on the rare-earth ions, which order approximately below 10 K. Here we emphasize that overall features of the phase diagram of Gd$_{1-x}$Tb$_x$MnO$_3$ can be reproduced within the Mn $3d$-spin sublattice model. However, to reproduce its subtle features, we may need to consider couplings between the Mn $3d$ spins and the rare-earth $f$-electron moments~\cite{Feyerherm06,Prokhnenko07a,Prokhnenko07b,Mukhin04,Hemberger04}. By contrast, in the case of Eu$_{1-x}$Y$_x$MnO$_3$ without interference from $f$-electron moments, the agreement between the experiment and the calculation is quite good.

To summarize this section, the $ab$- and $bc$-cycloidal spin states in $R$MnO$_3$ are stabilized by the single-ion anisotropy or the DM interaction. Namely, the hard magnetization $c$ axis due to the single-ion anisotropy gives a relative stability to the $ab$-cycloidal state, while the $bc$-cycloidal state is stabilized by the DM vectors on the out-of-plane bonds. On the other hand, the $ac$-plane spin cycloid is unfavorable. Recent first-principles calculations for TbMnO$_3$ also confirmed this tendency~\cite{HJXiang08,Malashevich08}.

\subsection{Magnetic structures in the cycloidal spin states}
\label{SSec:Ellipticity}
%%%%%%%%%%%%%%%%%%%%%%%%%%%%%%%%%%%%%%%%%%%%%%%%%%%%%%%%%%%%
\begin{figure}[tdp]
\includegraphics[scale=1.0]{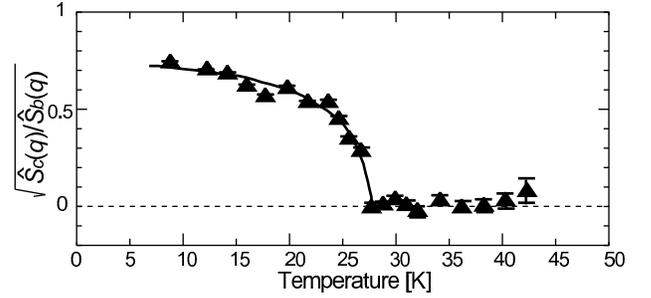}
\caption{Temperature dependence of $\sqrt{\hat{S}_c(\bm q)/\hat{S}_b(\bm q)}$ in TbMnO$_3$ measured in the spin-polarized neutron-scattering experiment from Ref.~\cite{Yamasaki07a}. Here $\bm q$ is the propagation wave vector of the $bc$-cycloidal spin state.}
\label{Fig11}
\end{figure}
%%%%%%%%%%%%%%%%%%%%%%%%%%%%%%%%%%%%%%%%%%%%%%%%%%%%%%%%%%%%
%%%%%%%%%%%%%%%%%%%%%%%%%%%%%%%%%%%%%%%%%%%%%%%%%%%%%%%%%%%%
\begin{figure*}[tdp]
\includegraphics[scale=1.0]{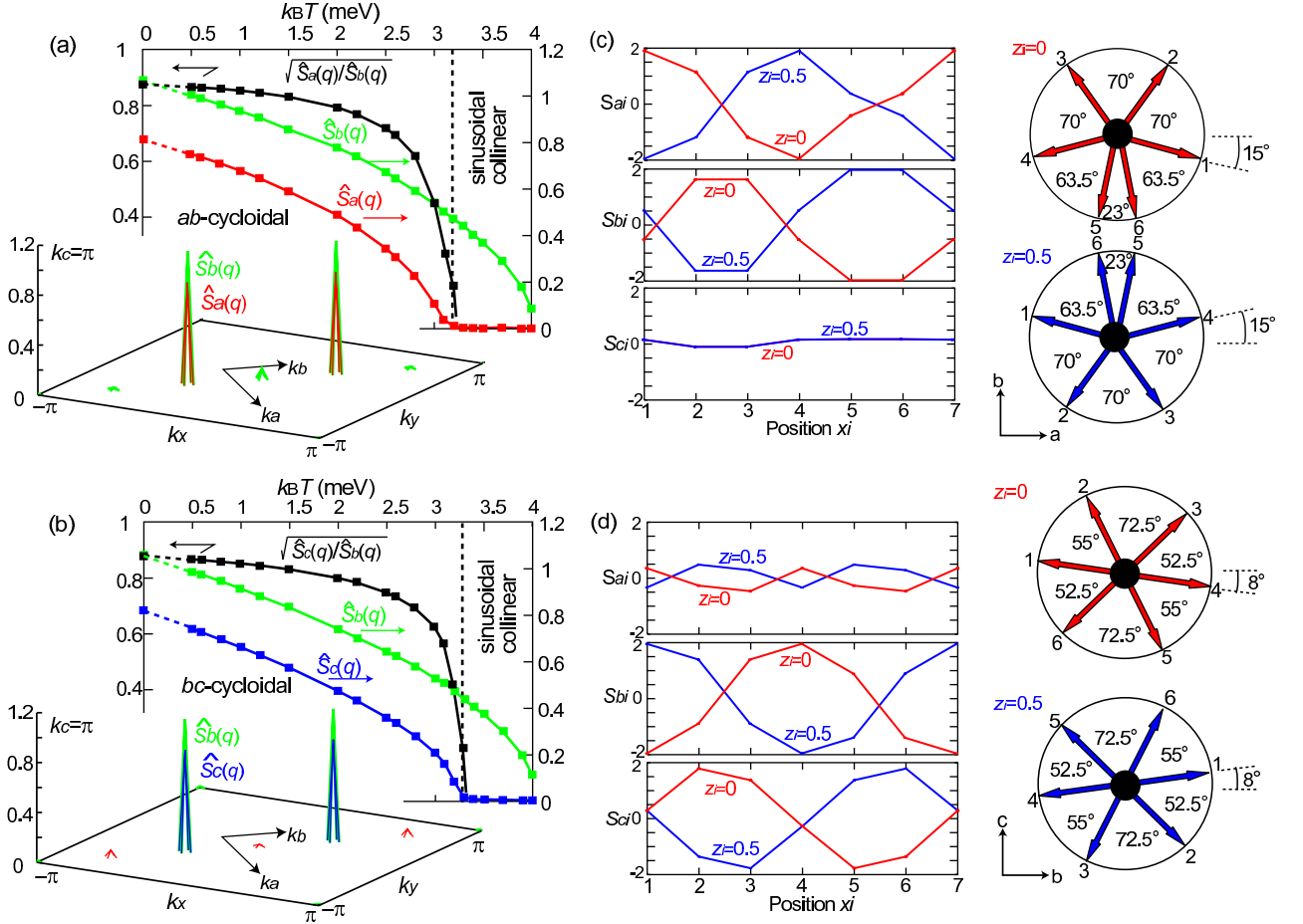}
\caption{(Color online) (a) Temperature dependence of $\hat{S}_a(\bm q)$, $\hat{S}_b(\bm q)$ and $\sqrt{\hat{S}_a(\bm q)/\hat{S}_b(\bm q)}$ for the $ab$-cycloidal spin state with $J_2$=~0.80 meV and $\alpha_c$=~0.24 meV. (b) Temperature dependence of $\hat{S}_b(\bm q)$, $\hat{S}_c(\bm q)$ and $\sqrt{\hat{S}_c(\bm q)/\hat{S}_b(\bm q)}$ for the $bc$-cycloidal spin state with $J_2$=~0.80 meV and $\alpha_c$=~0.38 meV. The data for finite temperatures obtained by the Monte-Carlo simulations are smoothly extrapolated to the data at $T$=~0 obtained by numerically solving the Landau-Lifshitz-Gilbert equation as indicated by dashed lines. Insets of (a) and (b) show the spin-correlation functions in the $k_x$-$k_y$ plane for $k_c=\pi$ at $T$=0. (c) Spin alignment (left panel) and spin directions (right panel) at $T$=0 for the $ab$-cycloidal spin state. Here the orthorhombic unit cell contains alternately stacked two different MnO planes with $z_i$=0 and 0.5. (d) Those for the $bc$-cycloidal spin state.}
\label{Fig12}
\end{figure*}
%%%%%%%%%%%%%%%%%%%%%%%%%%%%%%%%%%%%%%%%%%%%%%%%%%%%%%%%%%%%
The calculated spin correlation functions $\hat{S}_a(\bm k)$ and $\hat{S}_b(\bm k)$ in the $ab$-cycloidal spin phase are not equal in magnitude of the peak ----- see Fig.~\ref{Fig08}(d). This is also the case for $\hat{S}_b(\bm k)$ and $\hat{S}_c(\bm k)$ in the $bc$-cycloidal spin phase ----- see Fig.~\ref{Fig08}(c). At first sight, this seems to mean that the magnitude of ordered magnetic moment depends on its direction. The inequivalency between $\hat{S}_b(\bm q)$ and $\hat{S}_c(\bm q)$ was actually observed in the recent spin-polarized neutron scattering experiments for TbMnO$_3$ and Tb$_{1-x}$Dy$_x$MnO$_3$, and was ascribed to an elliptical modulation of the spin cycloid~\cite{Yamasaki07a,Arima06}. Figure~\ref{Fig11} shows the measured temperature dependence of $\sqrt{\hat{S}_c(\bm q)/\hat{S}_b(\bm q)}$ for the $bc$-plane spin cycloid in TbMnO$_3$ where $\bm q$ is the spiral propagation wave vector~\cite{Yamasaki07a}. Importantly, the data for finite temperatures are extrapolated not to unity but to $\sim$0.8 at $T=0$. This may imply that the spin cycloid is elliptically modulated even in the ground state possibly due to quantum fluctuations. However, our calculation reveals that the peaks of two spin correlation functions at $T$=0 are not equal even within our classical model [see insets of Figs.~\ref{Fig12}(a) and \ref{Fig12}(b)], in which the quantum fluctuations are not taken into account, indicating that the consideration of quantum fluctuations is not necessarily required to explain this experimental observation. In the following, we address this issue by investigating the detailed spin structures in the cycloidal spin states.

We first show the calculated temperature dependence of $\sqrt{\hat{S}_a(\bm q)/\hat{S}_b(\bm q)}$ for the $ab$-cycloidal case in Fig.~\ref{Fig12}(a), and that of $\sqrt{\hat{S}_c(\bm q)/\hat{S}_b(\bm q)}$ for the $bc$-cycloidal case in Fig.~\ref{Fig12}(b). They are calculated for $J_2$=~0.80 meV and $\alpha_c$=~0.24 meV, and for $J_2$=~0.80 meV and $\alpha_c$=~0.38 meV, respectively. Since the commensurate spin cycloid is convenient for discussion, we choose $J_2$=~0.80 meV (i.e. $J_{ab}/J_2$=~1), for which the spiral propagation wave number becomes $q_m=1/3$. This value corresponds to an intermediate value of $q_m\sim$~0.28 in TbMnO$_3$ and $q_m\sim$~0.36 in DyMnO$_3$. The data for finite temperatures are obtained by the Monte-Carlo simulations, while those for the ground states are obtained by numerically solving the Landau-Lifshitz-Gilbert equation. They are smoothly connected, and reproduce the experimentally observed temperature dependence and the non-unity values at $T$=0 for both $ab$- and $bc$-cycloidal cases.

To clarify the origin of non-unity values of $\sqrt{\hat{S}_a(\bm q)/\hat{S}_b(\bm q)}$ and $\sqrt{\hat{S}_c(\bm q)/\hat{S}_b(\bm q)}$ at $T$=0, we investigate detailed spin structures in the ground state for both $ab$- and $bc$-cycloidal cases. The results are shown in Figs.~\ref{Fig12}(c) and \ref{Fig12}(d). Noticeably, the rotation angles of cycloidal spins are not uniform, but suffer from considerable modulation. This rotation-angle modulation is due mainly to the single-ion spin anisotropies. The spins avoid to point along the hard magnetization axes, but prefer to direct along the easy magnetization axes. This is the very origin of the above inequivalent peak values of two kinds of spin correlation functions. We expect that higher harmonic peaks of the spin correlation functions should appear somewhere in the momentum space. In the $ab$-cycloidal case, we can see additional tiny peaks in the inset of Fig.~\ref{Fig12}(a).
%%Although the corresponding experiment for the $ab$-cycloidal spin phase 
%%does not exist presently, the predicted inequivalent $S_a(\bm q)$ and 
%%$S_b(\bm q)$ would also be observed in the spin-polarized neutron 
%%scattering experiment.

\subsection{Effects of single-ion anisotropies in the collinear spin states}
\label{SSec:SIAAFMA}
%%%%%%%%%%%%%%%%%%%%%%%%%%%%%%%%%%%%%%%%%%%%%%%%%%%%%%%%%%%%
\begin{figure}[tdp]
\includegraphics[scale=1.0]{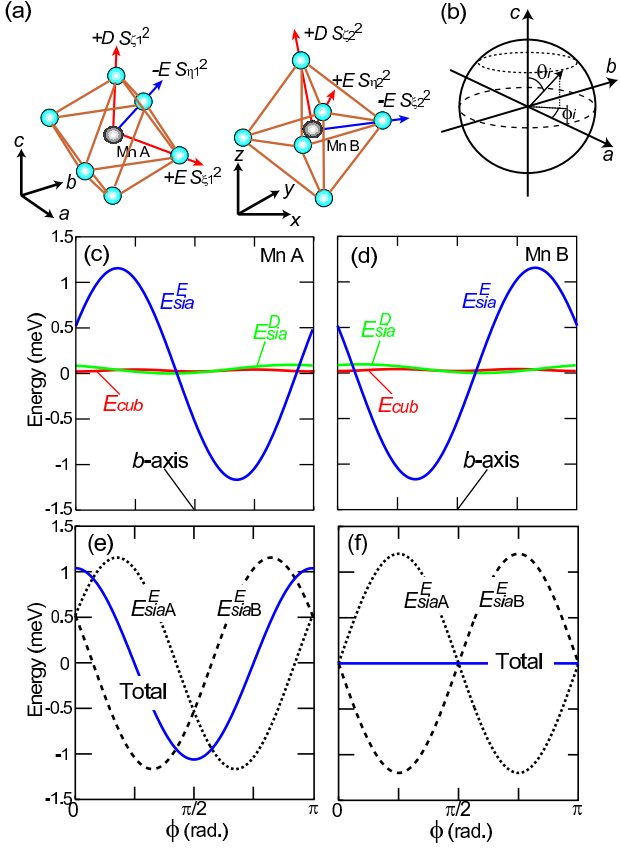}
\caption{(Color online) (a) Local hard magnetization axes (solid arrows) and local easy magnetization axes (dashed arrows) at Mn A and Mn B sites due to the single-ion anisotropies, $\mathcal{H}_{\rm sia}=\mathcal{H}_{\rm sia}^D+\mathcal{H}_{\rm sia}^E$. Owing to the tilting of MnO$_6$ octahedra, these axes deviate from the cubic $x$, $y$ and $z$ axes. (b) Polar representation of the $S$=2 classical-vector spin. The angle $\phi_i$ is defined with respect to the orthorhombic $a$ and $b$ axes. [(c) and (d)] $\phi_i$ dependence of energies $E_{\rm sia}^E$, $E_{\rm sia}^D$ and $E_{\rm cub}$ at (c) Mn A and (d) Mn B for a spin sticking on the $ab$ plane. [(e) and (f)] $\phi_i$ dependence of energies $E_{\rm siaA}^E$, $E_{\rm siaB}^E$ and $E_{\rm siaA}^E+E_{\rm siaB}^E$ in the (e) presence and (f) absence of the MnO$_6$ tilting. Here $E_{\rm siaA}^E$ and $E_{\rm siaB}^E$ are the energies of $\mathcal{H}_{\rm sia}^E$ at Mn A and Mn B sites, respectively. For the MnO$_6$ tilting, we use the structural parameters of EuMnO$_3$ taken from Ref.~\cite{Dabrowski05}.}
\label{Fig13}
\end{figure}
%%%%%%%%%%%%%%%%%%%%%%%%%%%%%%%%%%%%%%%%%%%%%%%%%%%%%%%%%%%%
In the (nearly) collinear magnetic phases in $R$MnO$_3$ like AFM(A) and sinusoidal phases, the Mn spins are aligned parallel to the $b$ axis. The issue which interaction or which magnetic anisotropy is responsible for this easy-axis spin anisotropy has not been clarified yet. In the Hamiltonian~(\ref{eq:model1}), there are some suspects, e.g. the single-ion anisotropy terms ($\mathcal{H}_{\rm sia}^D$ and $\mathcal{H}_{\rm sia}^E$), the DM-interaction term ($\mathcal{H}_{\rm DM}$), and the cubic-anisotropy term ($\mathcal{H}_{\rm cub}$). Among them, the DM-interaction term turns out to be irrelevant. We confirm that the spins in these phases are still parallel to the $b$ axis even without DM interaction.

In the literature, Matsumoto discussed a combination effect of cubic anisotropy $\mathcal{H}_{\rm cub}$ and one of the single-ion anisotropies $\mathcal{H}_{\rm sia}^D$ in the tilted MnO$_6$ octahedra~\cite{Matsumoto70}. Using the polar representation $(S_{ai}, S_{bi}, S_{ci})=(S\sin\theta_i\cos\phi_i, S\sin\theta_i\sin\phi_i, S\cos\theta_i)$ ----- see Fig.~\ref{Fig13}(a), and considering that the hard magnetization along the $c$ axis favors spins lying on the $ab$ plane (i.e. $\theta_i$=~$\pi$/2), the term $\mathcal{H}_{\rm cub}$ can be rewritten as
%%%%%%%%%%%%%%%%%%%%%%%%%%%%%%%%%%%
\begin{eqnarray}
\mathcal{H}_{\rm cub}
&\propto&\sum_{i}(S_{xi}^4+S_{yi}^4+S_{zi}^4) \nonumber \\
&=&S^4\sum_{i} (3-\cos4\phi_i)/4.
\label{eq:HcubPR}
\end{eqnarray}
%%%%%%%%%%%%%%%%%%%%%%%%%%%%%%%%%%%
This expression implies that the cubic anisotropy favors spins pointing along the $a$ or $b$ axis since this term has energy minima at $\phi_i$=~0, $\pi/2$, $\pi$, and $3\pi/2$. Furthermore, tilting of the MnO$_6$ octahedra makes the local $\zeta_i$ axes, which are local hard magnetization axes at every Mn site, inclined towards the $a$ axis, resulting in a relatively hard magnetization along the $a$ axis. Seemingly, this leads to a relatively easy magnetization along the $b$ axis. The energy of $\mathcal{H}_{\rm cub}+\mathcal{H}_{\rm sia}^D$ in the polar representation with $\theta_i=\pi/2$ indeed gives energy minima when the spin points in the $\pm b$ direction (i.e. $\phi_i$=~$\pi/2$ and $3\pi/2$).

However, this scenario turns out to be wrong since the energies of $\mathcal{H}_{\rm cub}$ and $\mathcal{H}_{\rm sia}^D$ ($E_{\rm cub}$ and $E_{\rm sia}^D$) are negligibly small as compared to the energy of $\mathcal{H}_{\rm sia}^E$ ($E_{\rm sia}^E$) when the spins stick to the $ab$ plane. Instead, the magnetic anisotropy is governed by the term $\mathcal{H}_{\rm sia}^E$. Because of the term $\mathcal{H}_{\rm sia}^E$, the $\xi_i$ and $\eta_i$ axes become a hard magnetization axis alternately in the $ab$ plane. In Figs.~\ref{Fig13}(c) and \ref{Fig13}(d), the $\phi_i$ dependence of $E_{\rm sia}^E$ is shown together with those of $E_{\rm cub}$ and $E_{\rm sia}^D$ for Mn A and Mn B sites, respectively. Here Mn A and Mn B are two different Mn sites in the same Mn-O plane ------ see Fig.~\ref{Fig01}(a). Noticeably, the energies $E_{\rm cub}$ and $E_{\rm sia}^D$ are much smaller than the energy $E_{\rm sia}^E$. However, the energy $E_{\rm sia}^E(\phi_i)$ has energy minima not along the $b$ axis ($\phi_i=\pi/2$), but nearly along the $x$ and $y$ axes ($\phi_i$=$\pi/4$ and $3\pi/4$), respectively. This may seem incompatible with the easy magnetization along the $b$ axis. However, since the in-plane ferromagnetic exchanges $J_{ab}$ predominantly force the spins at Mn A and Mn B sites to be parallel, a sum of these two energies, i.e. $E_{\rm sia}^E$ at Mn A and $E_{\rm sia}^E$ at Mn B ($E_{\rm siaA}^E$ and $E_{\rm siaB}^E$) governs the magnetic anisotropy in the collinear spin phases. We show the $\phi_i$ dependence of energies $E_{\rm siaA}^E$, $E_{\rm siaB}^E$ and $E_{\rm siaA+B}^E$=$E_{\rm siaA}^E$+$E_{\rm siaB}^E$ in Fig.~\ref{Fig13}(e). This figure actually shows that the energy minimum of $E_{\rm siaA+B}^E(\phi_i)$ is located at the $b$ axis ($\phi_i$=$\pi/2$).

In fact, this is a consequence of the MnO$_6$ tilting. Without tilting, the energy sum $E_{\rm siaA+B}^E(\phi_i)$ is always zero irrespective of $\phi_i$ ----- see Fig.~\ref{Fig13}(f). Although both $E_{\rm siaA}^E(\phi_i)$ and $E_{\rm siaB}^E(\phi_i)$ have strong $\phi_i$ dependence, they perfectly cancel out. The above discussion is valid also for Mn C and Mn D sites in another Mn-O plane because of the presence of a mirror plane between two Mn-O planes. We conclude that the combination effect of the single-ion anisotropies $\mathcal{H}_{\rm sia}=\mathcal{H}_{\rm sia}^D+\mathcal{H}_{\rm sia}^E$ and the GdFeO$_3$-type distortion is an origin of the easy-axis spin anisotropy in the AFM(A) and sinusoidal phases.

\subsection{Sinusoidal collinear spin phase}
\label{SSec:SinColl}
%%%%%%%%%%%%%%%%%%%%%%%%%%%%%%%%%%%%%%%%%%%%%%%%%%%%%%%%%%%%
\begin{figure}[tdp]
\includegraphics[scale=1.0]{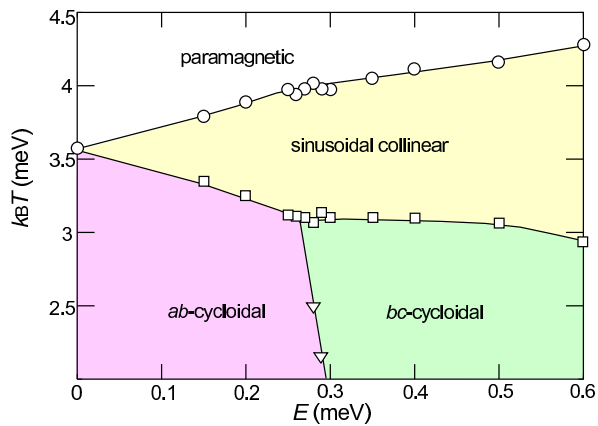}
\caption{$T$-$E$ phase diagram obtained by the Monte-Carlo analysis of the Hamiltonian (\ref{eq:model1}). Here the parameter $E$ is a variable which expresses the strength of single-ion anisotropy $\mathcal{H}_{\rm sia}^E$, and $J_2$ is fixed at 0.80 meV. Values of other model parameters in (\ref{eq:model1}) are fixed at the values listed in Table~\ref{tabl:MDLPRMS} (see Sec.~\ref{Sec:ModelMethod}). The temperature range over which the sinusoidal collinear spin phase emerges increases as $E$ is increased.}
\label{Fig14}
\end{figure}
%%%%%%%%%%%%%%%%%%%%%%%%%%%%%%%%%%%%%%%%%%%%%%%%%%%%%%%%%%%%
In this section, we discuss the nature and origin of the sinusoidal collinear spin phase in the intermediate temperature regime. In this phase, the spins are aligned along the $b$ axis with sinusoidally modulated amplitudes. We find that for its emergence, the single-ion anisotropy term $\mathcal{H}_{\rm sia}^E$ is essentially important. In Fig.~\ref{Fig14}, we display evolution of each magnetic phase as a function of the anisotropy parameter $E$. This figure shows that the temperature range over which the sinusoidal collinear phase emerges increases as $E$ is increased.

%%%%%%%%%%%%%%%%%%%%%%%%%%%%%%%%%%%%%%%%%%%%%%%%%%%%%%%%%%%%
\begin{figure}[tdp]
\includegraphics[scale=1.0]{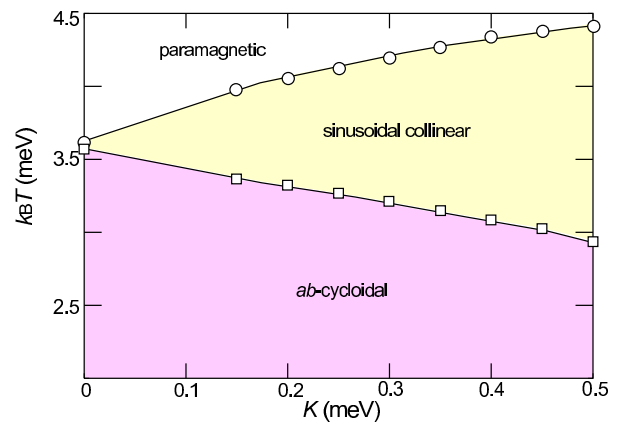}
\caption{$T$-$K$ phase diagram obtained by the Monte-Carlo calculation for the case without GdFeO$_3$-type distortion (see text). The term $\mathcal{H}_K$ given in Eq.(\ref{eq:SIAK}) is added to the Hamiltonian (\ref{eq:model1}), which generates the easy-axis spin anisotropy along the $b$ axis. Here, the parameter $K$ is a variable which expresses the strength of easy-axis spin anisotropy. Values of the model parameters except for $K$ and $J_2$=0.80 meV are fixed at the values listed in Table~\ref{tabl:MDLPRMS} (see Sec.~\ref{Sec:ModelMethod}). The temperature range of the sinusoidal collinear spin phase increases as $K$ is increased.}
\label{Fig15}
\end{figure}
%%%%%%%%%%%%%%%%%%%%%%%%%%%%%%%%%%%%%%%%%%%%%%%%%%%%%%%%%%%%
As we have discussed in the previous section, when the GdFeO$_3$-type distortion is present, the term $\mathcal{H}_{\rm sia}^E$ generates an easy-axis spin anisotropy along the $b$ axis in the collinear spin phase. In fact, this easy-axis anisotropy is decisively important for the stability of the sinusoidal collinear phase. To see the importance of easy-axis anisotropy, we have performed calculations for the cubic-symmetry case, i.e. the case without GdFeO$_3$-type distortion. We set $a$=$b$, $x_{{\rm O}_1}$=0, $y_{{\rm O}_1}$=0.5, $x_{{\rm O}_2}$=0.75, $y_{{\rm O}_2}$=0.25, and $z_{{\rm O}_2}$=0 for the local-axis vectors $\bm {\xi}_i$, $\bm {\eta}_i$ and $\bm {\zeta}_i$ in Eqs.~(\ref{eq:tilax1})-(\ref{eq:tilax3}). As a result, the local axes $\xi_i$, $\eta_i$ and $\zeta_i$ attached to the $i$-th MnO$_6$ are equivalent to the crystallographic $x$, $y$ and $z$ axes. In this case, the term $\mathcal{H}_{\rm sia}$ does not generate an easy-axis anisotropy, and the sinusoidal collinear phase indeed disappears. Instead, we further add a term of the easy-axis anisotropy $\mathcal{H}_K$ by hand to the Hamiltonian (\ref{eq:model1}), which is given by
%%%%%%%%%%%%%%%%%%%%%%%%%%%%%%%%%%%%%%%%%%%%%%%%%%
\begin{equation}
\mathcal{H}_K=-K\sum_{i}S_{b i}^2,
\label{eq:SIAK}
\end{equation}
%%%%%%%%%%%%%%%%%%%%%%%%%%%%%%%%%%%%%%%%%%%%%%%%%%
with $K>0$.
We obtain similar evolution of the sinusoidal collinear phase as a function of the parameter $K$ ----- see Fig.~\ref{Fig15}. Note that when $K$=0, the sinusoidal collinear phase vanishes, indicating that the GdFeO$_3$-type distortion or the easy-axis anisotropy is indispensable for its emergence. We indeed notice that in the experimental phase diagram of Fig.~\ref{Fig01}(b), the temperature range of the sinusoidal collinear phase is larger in a material with stronger GdFeO$_3$-type distortion, or a smaller $R$ ion.
%%In the $T$-$E$ diagram of Fig.~\ref{Fig14}, the sinusoidal collinear
%%phase does not vanish even at $E$=0 since a weak easy-axis anisotropy
%%still survives due to the term $\mathcal{H}_{\rm sia}^D$.)

The above facts can be understood easily as previously discussed in Ref.~\cite{Kenzelmann06}. As temperature is lowered, the first ordered state has spins aligned along the easy axis, i.e. the $b$ axis. As temperature is further lowered, the spin length grows. Then, the system develops a long-range transverse sinusoidal order to satisfy the competition between $J_{ab}$ and $J_2$. This requires to enter the cycloidal spin phase. The temperature range of the sinusoidal collinear phase increases as the anisotropy $K$ is increased because the longer spin length is required to overcome the larger $K$, which in turn requires going to lower temperature.

\section{Conclusion and Discussion}
\label{Sec:Conc-Disc}
In summary, we have studied the magnetoelectric phase diagrams of the perovskite manganites $R$MnO$_3$ by constructing a microscopic spin model to describe the Mn 3$d$-spin systems. We have analyzed this model by using the Monte-Carlo method for the thermodynamic properties and by numerically solving the Landau-Lifshitz-Gilbert equation for the ground-state properties. Considering that the GdFeO$_3$-type lattice distortion enhances the second-neighbor antiferromagnetic exchanges $J_2$, we have studied the $T$-$J_2$ phase diagrams, and have obtained diagrams in good agreement with the experimental ones including two kinds of multiferroic phases, i.e. the $ab$-cycloidal spin phase with $P_a$ and the $bc$-cycloidal spin phase with $P_c$.

We have discussed a mechanism of the electric polarization flop. The $ab$-cycloidal spin state is stabilized by the single-ion anisotropy $\mathcal{H}_{\rm sia}^D$ and the DM interaction $\mathcal{H}_{\rm DM}^{ab}$ with vectors on the in-plane bonds. Here the single-ion anisotropy $\mathcal{H}_{\rm sia}^D$ makes magnetization along the $c$ axis hard. On the other hand, the $bc$-cycloidal spin state is stabilized by the DM interaction $\mathcal{H}_{\rm DM}^c$ with vectors on the out-of-plane bonds. As the spiral rotation angle increases with increasing $J_2$, the energy gain due to $\mathcal{H}_{\rm DM}^{ab}$ is reduced in the $ab$-cycloidal spin state. In this way, the increasing GdFeO$_3$-type distortion destabilizes the $ab$-cycloidal spin state. This leads to the cycloidal-plane flop from $ab$ to $bc$, and consequently to the electric-polarization flop from $P_a$ to $P_c$.

We have demonstrated that the regime of $bc$-cycloidal spin phase with $P_c$ increases with increasing $\alpha_c$, and have revealed that the phase diagram of Gd$_{1-x}$Tb$_x$MnO$_3$ with a large $P_c$ regime is reproduced for a rather large value of $\alpha_c$=~0.38 meV, while that of Eu$_{1-x}$Y$_x$MnO$_3$ with a small $P_c$ regime is reproduced for a smaller value of $\alpha_c$=~0.30 meV. So far existence of the two different ferroelectric phases with $P_a$ and $P_c$ in $R$MnO$_3$ compounds have often been connected with the role of $f$-electron moments on the rare-earth ions. However our result indicates that the rare-earth magnetism plays a relatively minor role.

By investigating the detailed structures of the cycloidal spin states, we have found that the spiral rotation angles are not uniformly the same but are significantly distributed due to the single-ion anisotropies. The experimentally observed direction-dependent magnitude of spin-correlation peak is ascribed to this rotation-angle distribution instead of the elliptical modulation of the spin cycloid as claimed so far.

By examining the effect of single-ion anisotropies, we have found that because of the tilting of MnO$_6$ octahedra, the single-ion anisotropies, $\mathcal{H}_{\rm sia}=\mathcal{H}_{\rm sia}^D+\mathcal{H}_{\rm sia}^E$, energetically favor spins pointing along the $b$ axis in the AFM(A) and sinusoidal collinear phases.

We also found the sinusoidal collinear spin phase is stabilized by the easy-axis spin anisotropy along the $b$ axis generated by the single-ion anisotropy term $\mathcal{H}_{\rm sia}$ in the distorted lattice structure of GdFeO$_3$ type.

The microscopic model proposed here must be helpful for studying origins and mechanisms of several intriguing magnetoelectric phenomena discovered in the present manganite compounds, e.g., origins of the electrically activated spin excitation (electromagnon)~\cite{MiyaharaCD}, mechanisms of the magnetic-field-induced electric polarization flop~\cite{Aliouane09,Feyerherm09,Barath08}, and origins of the giant magnetocapacitance effect~\cite{Kagawa09,Schrettle09}. Discussion and analysis of the spin-wave dispersions obtained in the neutron-scattering experiments~\cite{Senff07,Senff08} are also an issue of interest.

Now, we would like to compare our work with the recent theoretical work done by Dagotto and coworkers, which is based on the two-orbital double-exchange model with additional terms~\cite{Sergienko06a,DongS08}. In Ref.~\cite{Sergienko06a}, they added the DM-interaction term with DM vectors coupling to the lattice, and reproduced the cycloidal spin order. On the other hand, in Ref.~\cite{DongS08}, they considered very weak second-neighbor spin exchanges to reproduce a successive emergence of the AFM(A), cycloidal, and AFM(E) phases, in the ground state. However, in those studies, they failed to reproduce the cycloidal-plane flop and the sinusoidal collinear phase. In addition, within their theory, it seems hard to explain the monotonical increase of the propagation wave number $q_m^{\rm Mn}$ (or the spiral rotation angle $\theta_s$) with increasing GdFeO$_3$-type distortion observed in experiments~\cite{Arima06,Yamasaki07b,Goto04}. On the other hand, we consider that the second-neighbor exchanges $J_2$ play a predominant role in stabilizing the cycloidal spin state, and the ferroelectric polarization subsequently emerges through the inverse DM mechanism. Our model in which the strength of $J_2$ scales with the orthorhombic lattice distortion naturally explains the observed monotonical increase of $q_m^{\rm Mn}$. Moreover, the spiral rotation angle determined by the ratio $J_{ab}$/$J_2$ turns out to control the competition between the $ab$- and $bc$-cycloidal spin states, resulting in the electric-polarization flop. This indicates that consideration of the second-neighbor exchanges $J_2$ is essentially important to describe the magnetoelectric coupling in the manganite system. Importance of the further exchanges is peculiar in the present manganites. This is  owing to the weak nearest-neighbor exchanges resulting from the cancellation of ferromagnetic and antiferromagnetic contributions from the $e_g$- and $t_{2g}$-orbital sectors due to the $t_{2g}^3e_g^1$ electron configuration. Indeed the values of $J_{ab}\sim$1.6 meV in LaMnO$_3$~\cite{Moussa96,Hirota96} is much smaller than $J_{ab}$=~13.3 meV in LaTiO$_3$~\cite{Keimer00}, which is a $S$=1/2 G-type antiferromagnet with $t_{2g}^1$ electron configuration.

%%%%%%%%%%%%%%%%%%%%%%%%%%%%%%%%%%%%%%%%%%%%%%%%%%%%%%%%%%%%
\begin{figure}[tdp]
\includegraphics[scale=1.0]{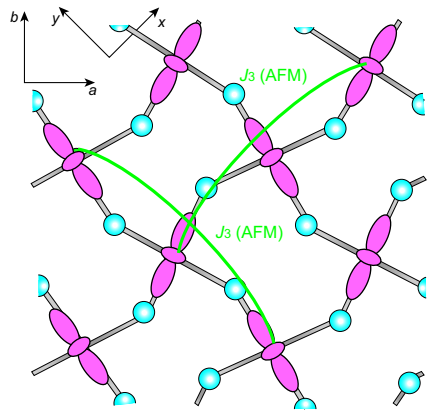}
\caption{(Color online) Third-neighbor spin exchange $J_3$ considered in Refs.~\cite{Solovyev08,Solovyev09}.}
\label{Fig16}
\end{figure}
%%%%%%%%%%%%%%%%%%%%%%%%%%%%%%%%%%%%%%%%%%%%%%%%%%%%%%%%%%%%%
It is also worthy mentioning that on the basis of the LDA+U calculation, Solovyev proposed that the third-neighbor spin exchange $J_3$ (see Fig.~\ref{Fig16}) is increased by the GdFeO$_3$-type distortion, and this exchange also plays an important role for stability of the spiral spin order~\cite{Solovyev08,Solovyev09}. In this case, the propagation wave vector of the magnetic spiral is determined by the frustration among three exchanges $J_{ab}$, $J_2$ and $J_3$, and its wave number $q_m$ is given by,
%%%%%%%%%%%%%%%%%%%%%%%%%%%%%%%%%%%%%%%%%%%%%%%%%%
\begin{equation}
q_m=\frac{1}{\pi} \cos^{-1}[\frac{J_{ab}}{2(J_2+J_3)}].
\label{eq:qJ1J2J3}
\end{equation}
%%%%%%%%%%%%%%%%%%%%%%%%%%%%%%%%%%%%%%%%%%%%%%%%%%
 To check whether the consideration of $J_3$ affects the results presented in this paper, we have examined the effects of $J_3$. We have found that once the wave number $q_m$ is determined, the same results are obtained.

%%%%% Acknowledgement %%%%%
\section*{Acknowledgment}
We are pleased to acknowledge enlightening discussions with Y. Tokura and N. Nagaosa. MM would like to acknowledge discussions with F. Kagawa, S. Miyahara, N. Kida, H. Murakawa, Y. Yamasaki, J.S. Lee, S. Ishiwata, Y. Takahashi, and I. Solovyev, and technical advices on the numerical simulations from Y. Motome and D. Tahara.
%%This work is supported in part by...

\appendix
\section{}
\label{AppendixA}

%%%%%%%%%%%%%%%%%%%%%%%%%%%%%%%%%%%%%%%%%%%%%%%%%%%%%%%%%%%%%%%%%%
\begin{table*}
\caption{Structural parameters for $R$MnO$_3$. Here $r_{ab}^l$ and $r_{ab}^s$ denote lengths of the longer and shorter in-plane Mn-O bonds, respectively. References from which the data are taken are shown in the second column.}
\begin{tabular}{cc|ccccccccccccc}
\hline
$R$MnO$_3$ & Ref. & IR($\AA$) & $a$($\AA$) & $b$($\AA$) & $c$($\AA$) 
& $x_{\rm O_2}$ & $y_{\rm O_2}$ & $z_{\rm O_2}$ & $\phi$(deg) 
& $\varphi_c$(deg) & $\varphi_{ab}$(deg) & $r_c$($\AA$) & $r_{ab}^l$($\AA$) 
& $r_{ab}^s$($\AA$)\\
\hline
LaMnO$_3$ & (\cite{RodriguezC98}) & 1.216 & 5.5367 & 5.7473 & 7.6929 & 0.7256 & 0.3066 & 0.0384 & 107.06 & 155.48 &  155.11 & 1.9680 &  2.178 & 1.907\\
LaMnO$_3$ & (\cite{Dabrowski05}) & 1.216 & 5.5405 & 5.7458 & 7.6998 & 0.7261 & 0.3069 & 0.0388 & 106.43 & 155.28 &  155.04 & 1.9706 &  2.184 & 1.904\\

PrMnO$_3$ & (\cite{Alonso00}) & 1.179 & 5.4491 & 5.8129 & 7.5856 & 0.7151 & 0.3174 & 0.0430 & 110.93 & 152.36 &  150.51 & 1.9530 & 2.210 & 1.909\\
PrMnO$_3$ & (\cite{Dabrowski05}) & 1.179 & 5.4500 & 5.8295 & 7.5805 & 0.7151 & 0.3179 & 0.0433 & 111.41 & 151.59 &  150.36 & 1.9549 & 2.217 & 1.910\\

NdMnO$_3$ & (\cite{Alonso00}) & 1.163 & 5.4170 & 5.8317 & 7.5546 & 0.7141 & 0.3188 & 0.0450 & 111.52 & 150.86 &  149.63 & 1.9514 & 2.218 & 1.905\\
NdMnO$_3$ & (\cite{Mori02})   & 1.163 & 5.416  & 5.849  & 7.543  & 0.7132 & 0.3204 & 0.0450 & 112.09 & 150.3  &  149.2  & 1.951  & 2.227 & 1.905\\
NdMnO$_3$ & (\cite{Dabrowski05}) & 1.163 & 5.4168 & 5.8518 & 7.5479 & 0.7124 & 0.3199 & 0.0447 & 112.67 & 150.38 &  149.27 & 1.9520 & 2.223 & 1.911\\

SmMnO$_3$ & (\cite{Mori02})   & 1.132 & 5.362  & 5.862  & 7.477  & 0.7076 & 0.3241 & 0.0485 & 114.37 & 147.6  & 147.0   & 1.947  & 2.232 & 1.910\\

EuMnO$_3$ & (\cite{Mori02})   & 1.120    & 5.340  & 5.866  & 7.448  & 0.7065 & 0.3254 & 0.0487 & 114.82 & 146.54 &  146.47 & 1.944  &2.234  & 1.907\\
EuMnO$_3$ & (\cite{Dabrowski05}) & 1.120 & 5.3437 & 5.8361 & 7.4619 & 0.7055 & 0.3247 & 0.0485 & 114 55 & 147.35 &  146.45 & 1.9438 & 2.220 & 1.912\\

GdMnO$_3$ & (\cite{Mori02})   & 1.107 & 5.318  & 5.866  & 7.431  & 0.7057 & 0.3246 & 0.0508 & 115.74 & 145.6  & 146.0  & 1.944 & 2.229 & 1.915\\

TbMnO$_3$ & (\cite{Alonso00}) & 1.095 & 5.2931 & 5.8384 & 7.4025 & 0.7039 & 0.3262 & 0.0510 & 115.47 & 145.06 &  145.36 & 1.9401 & 2.221 & 1.905\\

DyMnO$_3$ & (\cite{Alonso00}) & 1.083 & 5.2785 & 5.8337 & 7.3778 & 0.7028 & 0.3276 & 0.0521 & 115.69 & 143.23 &  144.70 & 1.9437 & 2.224 & 1.903\\
DyMnO$_3$ & (\cite{Dabrowski05}) & 1.083 & 5.2802 & 5.8448 & 7.3789 & 0.7013 & 0.3249 & 0.0518 & 118.86 & 145.37 &  145.03 & 1.9322 & 2.210 & 1.919\\

YMnO$_3$ &  (\cite{Alonso00}) & 1.075 & 5.2418 & 5.8029 & 7.3643 & 0.7005 & 0.3266 & 0.0520 & 116.17 & 143.51 &  144.55 & 1.9385 & 2.200 & 1.904\\

HoMnO$_3$ & (\cite{Alonso00}) & 1.072 & 5.2572 & 5.8354 & 7.3606 & 0.7013 & 0.3281 & 0.0534 & 116.28 & 142.47 &  144.08 & 1.9435 & 2.222 & 1.905\\

ErMnO$_3$ & (\cite{Alonso00}) & 1.062 & 5.2262 & 5.7932 & 7.3486 & 0.7003 & 0.3266 & 0.0542 & 116.18 & 142.82 &  143.92 & 1.9382 & 2.199 & 1.903\\
ErMnO$_3$ & (\cite{Tachibana07}) & 1.062 & 5.2395 & 5.8223 & 7.3357 & 0.6997 & 0.3343 & 0.0563 & 113.59 & 141.59 &  142.05 & 1.942 &  2.248 & 1.891\\

TmMnO$_3$ & (\cite{Tachibana07}) & 1.052 & 5.2277 & 5.8085 & 7.3175 & 0.6984 & 0.3372 & 0.0571 & 112.75 & 141.15 &  141.17 & 1.940 &  2.255 & 1.886\\

YbMnO$_3$ & (\cite{Tachibana07}) & 1.042 & 5.2163 & 5.7991 & 7.2992 & 0.6979 & 0.3394 & 0.0580 & 112.07 & 140.27 &  140.51 & 1.940 &  2.263 & 1.879\\

LuMnO$_3$ & (\cite{Tachibana07}) & 1.032 & 5.1972 & 5.7868 & 7.2959 & 0.6989 & 0.3415 & 0.0575 & 110.14 & 139.56 &  140.36 & 1.944 &  2.269 & 1.862\\
\hline
\end{tabular}
\label{tabl:STPAPPNDX}
\end{table*}
%%%%%%%%%%%%%%%%%%%%%%%%%%%%%%%%%%%%%%%%%%%%%%%%%%%%%%%%%%%%%%%%%%
In this Appendix, we calculate the values of superexchange parameters, $J_{ab}$ and $J_c$, for $R$MnO$_3$ by following the formulation given by Gontchar and coworkers~\cite{Gontchar01,Gontchar02}.

In $R$MnO$_3$, the ground-state orbital state of each Mn$^{3+}$ ion is dominantly determined by local coordinations of surrounding oxygens with the Jahn-Teller distortion because of the strong electron-lattice coupling. The wave functions of occupied and hole orbitals at the $n$-th Mn$^{3+}$ site, $\Psi_{1n}$ and $\Psi_{2n}$, can be written as linear combinations of the twofold $e_g$-orbital wave functions~\cite{Kanamori60};
%%%%%%%%%%%%%%%%%%%%%%%%%%%%%%%%%%%%%%%%%%%%%%%%%%
\begin{eqnarray}
\Psi_{1n}&=&
 \sin\frac{\phi_n}{2}\varphi_{n\theta}
+\cos\frac{\phi_n}{2}\varphi_{n\eps},\\
\Psi_{2n}&=&
 \cos\frac{\phi_n}{2}\varphi_{n\theta}
-\sin\frac{\phi_n}{2}\varphi_{n\eps},
\end{eqnarray}
%%%%%%%%%%%%%%%%%%%%%%%%%%%%%%%%%%%%%%%%%%%%%%%%%%
with
%%%%%%%%%%%%%%%%%%%%%%%%%%%%%%%%%%%
\begin{eqnarray}
\varphi_{n \theta}&=&\frac{1}{2}(3\zeta_n^2-r^2), \\
\varphi_{n \eps}&=&\frac{\sqrt{3}}{2}(\xi_n^2-\eta_n^2),
\end{eqnarray}
%%%%%%%%%%%%%%%%%%%%%%%%%%%%%%%%%%%
where $\xi_n$, $\eta_n$ and $\zeta_n$ are coordinates with respect to the local axes attached to each MnO$_6$ octahedron.

The values of orbital angle $\phi_n$ can be derived from experimental data, using parameters of lattice distortions;
%%%%%%%%%%%%%%%%%%%%%%%%%%%%%%%%%%%
\begin{eqnarray}
\cos{\phi_n}&=&\frac{Q_{\theta n}}{\sqrt{Q_{\theta n}^2+Q_{\eps n}^2}},\\
\sin{\phi_n}&=&\frac{Q_{\eps n}}{\sqrt{Q_{\theta n}^2+Q_{\eps n}^2}},
\end{eqnarray}
%%%%%%%%%%%%%%%%%%%%%%%%%%%%%%%%%%%
with $Q_{\theta n}$ and $Q_{\eps n}$ being symmetrized Jahn-Teller distortions on each MnO$_6$ octahedron of $3z^2-r^2$ and $x^2-y^2$ types, respectively.

In the orthorhombic $R$MnO$_3$ crystal described by the $P_{bnm}$ space group, the Jahn-Teller distortions $Q_{\theta n}$ and $Q_{\eps n}$ are given by
%%%%%%%%%%%%%%%%%%%%%%%%%%%%%%%%%%%
\begin{eqnarray}
Q_{\theta}&=&\frac{1}{\sqrt{12}} [c - \frac{1}{\sqrt{2}}(a+b)],\\
Q_{\eps}&=&\sqrt{2} (v_x a+v_y b),
\end{eqnarray}
%%%%%%%%%%%%%%%%%%%%%%%%%%%%%%%%%%%
with
%%%%%%%%%%%%%%%%%%%%%%%%%%%%%%%%%%%
\begin{eqnarray}
v_x&=&\frac{3}{4}-x_{{\rm O}_2},\\
v_y&=&\frac{1}{4}-y_{{\rm O}_2}.
\end{eqnarray}
%%%%%%%%%%%%%%%%%%%%%%%%%%%%%%%%%%%
Here, $v_x$ and $v_y$ are shifts of the in-plane oxygen environment, $x_{{\rm O}_2}$ and $y_{{\rm O}_2}$ are coordination parameters for the in-plane oxygen ions, and $a$, $b$ and $c$ are the lattice parameters in $P_{bnm}$ axes ----- see Table.~\ref{tabl:STPAPPNDX}. 

The Jahn-Teller distortion in $R$MnO$_3$ causes a C-type orbital ordering, for which the orbital angles satisfy the following relation:
%%%%%%%%%%%%%%%%%%%%%%%%%%%%%%%%%%%%%%%%%%%%%%%%%%
\begin{equation}
\phi_1=\phi_2=-\phi_3=-\phi_4=\phi.
\end{equation}
%%%%%%%%%%%%%%%%%%%%%%%%%%%%%%%%%%%%%%%%%%%%%%%%%%

Within this framework, the superexchange parameters depend on the lattice distortions and the orbital structure as
%%%%%%%%%%%%%%%%%%%%%%%%%%%%%%%%%%%
\begin{eqnarray}
J_{ab}&=&\frac{J_0 \cos^2\varphi_{ab}}{r_{ab}^{10}}
[1 - \alpha \cos\phi + \beta(\cos^2 \phi+\frac{3}{4})],\\
J_{c}&=&\frac{J_0 \cos^2\varphi_{c}}{r_{c}^{10}}
[1 + 2\alpha \cos\phi - \beta \cos^2\phi],
\end{eqnarray}
%%%%%%%%%%%%%%%%%%%%%%%%%%%%%%%%%%%
where $\varphi_{ab}$ and $\varphi_{c}$ are means for the in-plane and out-of-plane Mn-O-Mn bond angles, respectively, and $r_{ab}$ and $r_{c}$ are means for the in-plane and out-of-plane Mn-O bond lengths, respectively.
For the values of parameters appeared in these formulas, we use $J_0$=1.456 eV$\AA$, $\alpha$=1.0 and $\beta$=4.5 according to Refs.~\cite{Gontchar01,Gontchar02}.

\begin{figure}[tdp]
\includegraphics[scale=1.0]{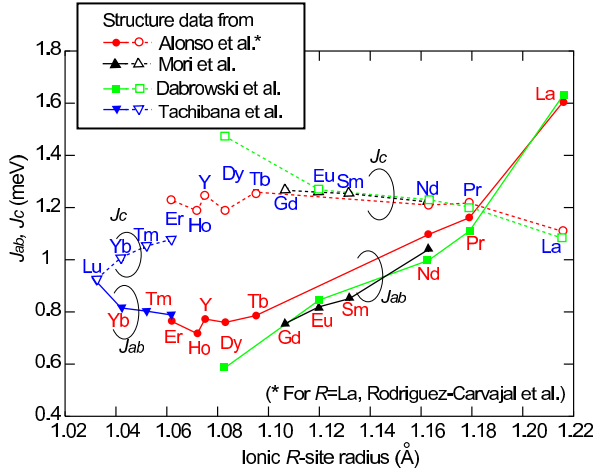}
\caption{(Color online) Calculated superexchange parameters, $J_{ab}$ and $J_c$, plotted as functions of the ionic $R$-site radius. Structural data used in the calculation are taken from Refs.~\cite{RodriguezC98,Alonso00,Mori02,Dabrowski05,Tachibana07}.}
\label{Fig17}
\end{figure}

%%%%%%%%%%%%%%%%%%%%%%%%%%%%%%%%%%%%%%%%%%%%%%%%%%%%%%%%%%%%%%%%%%
%%\begin{table}
%%\caption{Calculated values of superexchange parameters 
%%$J_{ab}$ and $J_c$ and those of the single-ion anisotropy 
%%parameters $D$ and $E$. 
%%Numbers near the compound names denote references from which 
%%the data are taken.}
%%\begin{tabular}{cc|cccc}
%%\hline
%%$R$MnO$_3$ & Ref. & $J_{ab}$(meV) & $J_c$(meV) & $D$(meV) & $E$(meV) \\
%%\hline
%%LaMnO$_3$ & (Ref.citex) & 1.60 & 1.11 & 0.17 & 0.26\\
%%LaMnO$_3$ & (Ref.citex) & 1.63 & 1.08 & 0.16 & 0.26\\
%%PrMnO$_3$ & (Ref.citex) & 1.16 & 1.22 & 0.20 & 0.25\\
%%PrMnO$_3$ & (Ref.citex) & 1.11 & 1.20 & 0.21 & 0.25\\
%%NdMnO$_3$ & (Ref.citex) & 1.10 & 1.21 & 0.21 & 0.25\\
%%NdMnO$_3$ & (Ref.citex) & 1.00 & 1.23 & 0.22 & 0.25\\
%%EuMnO$_3$ & (Ref.citex) & 0.85 & 1.27 & 0.24 & 0.25\\
%%TbMnO$_3$ & (Ref.citex) & 0.79 & 1.26 & 0.25 & 0.24\\
%%DyMnO$_3$ & (Ref.citex) & 0.76 & 1.19 & 0.25 & 0.24\\
%%DyMnO$_3$ & (Ref.citex) & 0.59 & 1.47 & 0.28 & 0.24\\
%%YMnO$_3$  & (Ref.citex) & 0.77 & 1.25 & 0.25 & 0.24\\
%%HoMnO$_3$ & (Ref.citex) & 0.72 & 1.19 & 0.25 & 0.24\\
%%ErMnO$_3$ & (Ref.citex) & 0.76 & 1.23 & 0.25 & 0.24\\
%%ErMnO$_3$ & (Ref.citex) & 0.76 & 1.08 & 0.23 & 0.25\\
%%TmMnO$_3$ & (Ref.citex) & 0.80 & 1.05 & 0.22 & 0.25\\
%%YbMnO$_3$ & (Ref.citex) & 0.82 & 1.01 & 0.21 & 0.25\\
%%LuMnO$_3$ & (Ref.citex) & 0.92 & 0.92 & 0.20 & 0.25\\
%%\hline
%%\end{tabular}
%%\label{tabl:MDLPRM}
%%\end{table}
%%%%%%%%%%%%%%%%%%%%%%%%%%%%%%%%%%%%%%%%%%%%%%%%%%%%%%%%%%%%%%%%%%
The values of $J_{ab}$ and $J_c$ calculated using structural data in Refs.~\cite{RodriguezC98,Alonso00,Mori02,Dabrowski05,Tachibana07} are plotted in Fig.~\ref{Fig17} as functions of the ionic $R$-site radius. Here, the values for LaMnO$_3$ were obtained from the spin-wave dispersion measured in the neutron-scattering experiments as $J_{ab}$=~1.66 meV and $J_c$=~1.16 meV in Ref.~\cite{Moussa96}, and $J_{ab}$=~1.67 meV and $J_c$=~1.21 meV in Ref.~\cite{Hirota96}. The values for PrMnO$_3$ were also obtained in the neutron-scattering experiment as $J_{ab}$=~1.12-1.19 meV and $J_c$=~1.2-1.29 meV in Ref.~\cite{Kajimoto05}.
The calculated values for LaMnO$_3$ and PrMnO$_3$ are in good agreement with these experimental values.
%%They are also listed in Table.~\ref{MDLPRM}.
We note that both $J_{ab}$ and $J_c$ are nearly constant in the region of relatively small-sized  $R$ ions, e.g. $R$=Eu, Tb, Dy and Y, in/near which the multiferroic phases emerge.
%%%%%%%%%%%%%%%%%%%%%%%%%%%%%%%%%%%%%%%%%%%%%%%%%%%%%%%%%%%%
%%%%%%%%%%%%%%%%%%%%%%%%%%%%%%%%%%%%%%%%%%%%%%%%%%%%%%%%%%%%%

\section{}
\label{AppendixB}

%%%%%%%%%%%%%%%%%%%%%%%%%%%%%%%%%%%%%%%%%%%%%%%%%%%%%%%%%%%%
\begin{figure}[tdp]
\includegraphics[scale=1.0]{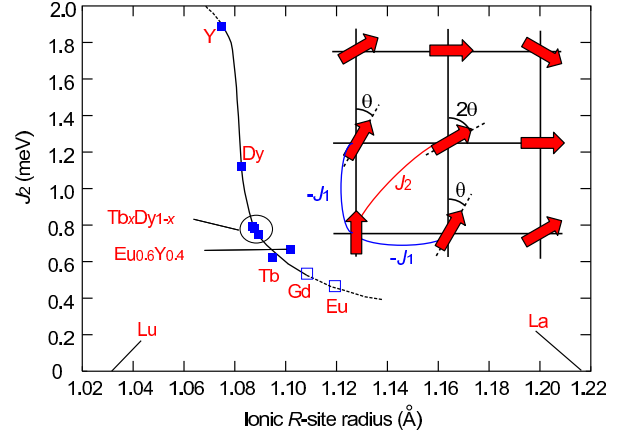}
\caption{(Color online) Estimated values of $J_2$ in $R$MnO$_3$ and several solid solutions plotted as functions of the ionic $R$-site radius. Inset describes the $J_1$-$J_2$ classical Heisenberg model on the anisotropic triangular lattice and the spiral spin order with uniform rotation angles of $\theta$.}
\label{Fig18}
\end{figure}
%%%%%%%%%%%%%%%%%%%%%%%%%%%%%%%%%%%%%%%%%%%%%%%%%%%%%%%%%%%%%
In this Appendix, we estimate the values of parameter $J_2$ for $R$MnO$_3$.
The spiral spin orders in $R$MnO$_3$ have turned out to be realized due to the competition between the nearest-neighbor ferromagnetic coupling $J_{ab}$ and the second-neighbor antiferromagnetic coupling $J_2$ in the $ab$ plane. Along the $c$ axis, the spins are forced to be aligned antiferromagnetically because of the strong antiferromagnetic coupling $J_c$. Thus the propagation wave number and the rotation angle of the magnetic spiral are dominantly determined by the ratio $J_2/J_{ab}$ although magnetic anisotropies and DM interactions should modulate them to some extent. Under this circumstance, we can roughly estimate the values of $J_2$ with the aid of the simple $J_1$-$J_2$ classical Heisenberg model on the anisotropic triangular lattice (see inset of Fig.~\ref{Fig18}). The Hamiltonian is given by
%%%%%%%%%%%%%%%%%%%%%%%%%%%%%%%%%%%%%%%%%%%%%%%%%%%%%%%%%%%%%
\begin{eqnarray}
\mathcal{H}_{J_1J_2}&=&
-J_1\sum_{<i,j>}^{n.n.}\bm S_i \cdot \bm S_j
+J_2\sum_{\ll i,j \gg}^{n.n.n.}\bm S_i \cdot \bm S_j \nonumber \\
&+&K \sum_{i}S_{zi}^2.
\end{eqnarray}
%%%%%%%%%%%%%%%%%%%%%%%%%%%%%%%%%%%%%%%%%%%%%%%%%%%%%%%%%%%%%
Here, $J_1$ and $J_2$ are the ferromagnetic exchange for the nearest-neighbor bonds and the antiferromagnetic exchange for the second-neighbor (diagonal) bonds, respectively. We introduce the third term with $K>0$ to determine the basal spiral plane by making the magnetization along the $z$ axis hard.

Within this model, the energy for spiral spin state per site is given as a function of the spiral rotation angle $\theta$;
%%%%%%%%%%%%%%%%%%%%%%%%%%%%%%%%%%%%%%%%%%%%%%%%%%%%%%%%%%%%%
\begin{equation}
E(\theta)/S^2=-2J_1\cos\theta + J_2\cos(2\theta).
\end{equation}
%%%%%%%%%%%%%%%%%%%%%%%%%%%%%%%%%%%%%%%%%%%%%%%%%%%%%%%%%%%%%
From a saddle-point equation, $dE(\theta)/d\theta=0$, we obtain a relation $\cos\theta_s=J_1/(2J_2)$ where $\theta_s$ is the rotation angle of spiral spin state with a minimum energy. This formula implies that a spiral spin state emerges when $J_2/J_1>0.5$, while when $J_2/J_1<0.5$ a ferromagnetic state is stabilized.

%%%%%%%%%%%%%%%%%%%%%%%%%%%%%%%%%%%%%%%%%%%%%%%%%%%%%%%%%%%%%%%%%%
\begin{table*}
\caption{On the basis of the simple two-dimensional $J_1$-$J_2$ classical Heisenberg model, the values of $J_2$ in TbMnO$_3$, DyMnO$_3$, YMnO$_3$ and several solid solutions are estimated from the experimental data of the spiral wave numbers $q_m^{\rm Mn}$ and theoretically calculated values of $J_{ab}$. Here, IR denotes the ionic radius of the $R$ ion, and $\theta$ is the spiral rotation angle calculated from $q_m^{\rm Mn}$ as $\theta=180^{\circ} \times q_m^{\rm Mn}$. The references from which the data of $q_m^{\rm Mn}$ are taken are presented in the last column. The effective $R$-site radii in the solid solutions are deduced by interpolation.}
\begin{tabular}{c|ccccccc}
\hline
 $R$MnO$_3$ & IR($\AA$) & 2$q_m^{\rm Mn}$ & $\theta$ (deg) & $J_1$ (meV) & 
 $\gamma=(2\cos\theta)^{-1}$ & $J_2=\gamma J_1$ (meV) & Ref. \\
\hline
TbMnO$_3$                   
& 1.095 & 0.56  & 50.4 & 0.79  & 0.784 & 0.62 &  \cite{Goto04} \\
DyMnO$_3$                   
& 1.083 & 0.78  & 70.2 & 0.76  & 1.47  & 1.12 &  \cite{Goto04} \\
YMnO$_3$                   
& 1.075 & 0.87  & 78.3 & 0.77  & 2.466 & 1.90 &  \cite{Munoz02} \\
Eu$_{0.6}$Y$_{0.4}$MnO$_3$  
& 1.102 & 0.58  & 52.2 & 0.817 & 0.815 & 0.67 &  \cite{Yamasaki07b} \\
Tb$_{0.32}$Dy$_{0.68}$MnO$_3$ 
& 1.087 & 0.678 & 61.0 & 0.768 & 1.032 & 0.79 &  \cite{Arima06} \\
Tb$_{0.41}$Dy$_{0.59}$MnO$_3$ 
& 1.088 & 0.663 & 59.6 & 0.786 & 0.990 & 0.78 &  \cite{Arima06} \\
Tb$_{0.50}$Dy$_{0.50}$MnO$_3$ 
& 1.089 & 0.656 & 59.0 & 0.773 & 0.972 & 0.75 &  \cite{Arima06} \\
\hline
\end{tabular}
\label{tabl:J2estm}
\end{table*}
%%%%%%%%%%%%%%%%%%%%%%%%%%%%%%%%%%%%%%%%%%%%%%%%%%%%%%%%%%%%%%%%%%
We estimate the values of $J_2$ in TbMnO$_3$, DyMnO$_3$, YMnO$_3$ and several solid solutions with the use of experimentally measured spiral wave numbers $q_m^{\rm Mn}$~\cite{Arima06,Yamasaki07b,Goto04} and the values of $J_1$ ($J_{ab}$) calculated in Appendix A (see Table~\ref{tabl:J2estm}). In Fig.~\ref{Fig18}, we plot them as a function of the ionic $R$-site radius, which exhibits a monotonical increase with decreasing $R$-site radius. We can deduce the values even for $R$MnO$_3$ with $R$=Eu and Gd by extrapolating the data, although their ground-state magnetic structure is not spiral but AFM(A).

When $q_m^{\rm Mn}$ is small, the energy difference between the spiral spin state and the AFM(A) state would be so small that the DM interaction and the single-ion anisotropy can affect their energetics significantly. Both of these two favor the AFM(A) state. The collinearly aligned spins parallel to the $b$ axis in the AFM(A) state can benefit from both the easy magnetization $b$ axis due to the single-ion anisotropy and the DM vectors on the out-of-plane bonds with a large $a$ component $\alpha_c$. Therefore, the estimation of $J_2$ based on the pure $J_1$-$J_2$ model performed here would involve ambiguities particularly near the AFM(A)-cycloidal phase boundary or in TbMnO$_3$. The actual value of $J_2$ in TbMnO$_3$ is expected to be slightly larger than the above-calculated value of $J_2=$~0.62 meV.

\section{}
\label{AppendixC}

%%%%%%%%%%%%%%%%%%%%%%%%%%%%%%%%%%%%%%%%%%%%%%%%%%%%%%%%%%%%
\begin{figure}[tdp]
\includegraphics[scale=1.0]{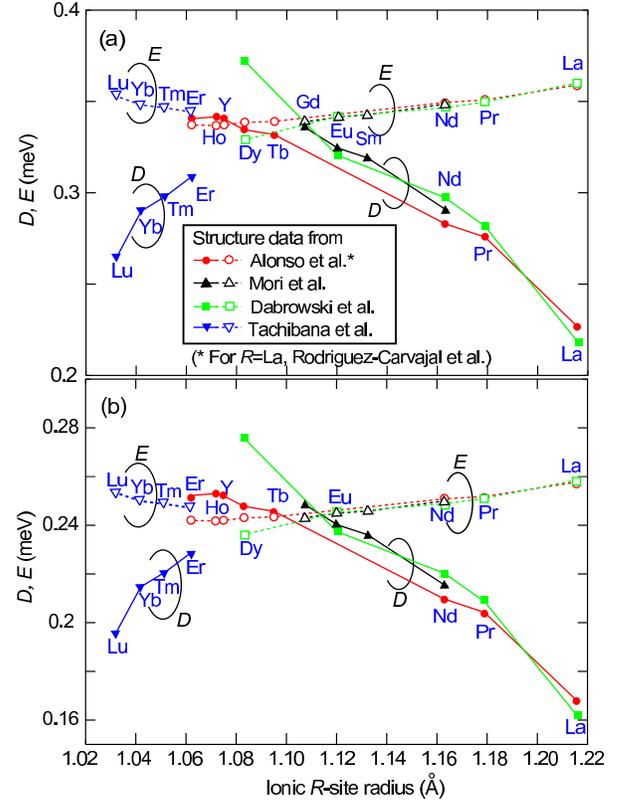}
\caption{(Color online) Calculated single-ion anisotropy parameters, $D$ and $E$ plotted as functions of the ionic $R$-site radius. Structural data used in the calculation are taken from Refs.~\cite{RodriguezC98,Alonso00,Mori02,Dabrowski05,Tachibana07}. Optical absorption data are taken from different references between (a) and (b), i.e. Ref.~\cite{Matsumoto70} for (a) and Ref.~\cite{Gerristen63} for (b).}
\label{Fig19}
\end{figure}
%%%%%%%%%%%%%%%%%%%%%%%%%%%%%%%%%%%%%%%%%%%%%%%%%%%%%%%%%%%%%
In this Appendix, we calculate the values of single-ion anisotropy parameters $D$ and $E$ in $R$MnO$_3$ with the use of formulas proposed by Matsumoto~\cite{Matsumoto70}.

The parameters $D$ and $E$ can be calculated within the second-order perturbation theory in terms of the spin-orbit interaction. They are given by
%%%%%%%%%%%%%%%%%%%%%%%%%%%%%%%%%%%
\begin{eqnarray}
D&=&-3(\frac{\lambda^2}{\Delta E}+\frac{4\lambda^2}{3\Delta})\cos\phi_{n},\\
E&=&-\frac{\sqrt{3}\lambda^2}{\Delta E}\sin\phi_{n},
\end{eqnarray}
%%%%%%%%%%%%%%%%%%%%%%%%%%%%%%%%%%%
where $\lambda$ is the coupling constant of the spin-orbit interaction, and $\phi_{n}$ is the orbital angle which represents the orbital wave functions at the $n$-th Mn$^{3+}$ site as defined in Appendix A. Here, $\Delta E$ and $\Delta$ are energy levels for the two excited states of Mn$^{3+}$ ion, $^3T_{1g}$ ($d\eps^4$) and $^5T_{2g}$ ($d\eps^2d\gamma^2$), measured from the ground-state $^5E_g$ ($d\eps^3d\gamma$) level, and satisfy the following relation~\cite{Sugano70};
%%%%%%%%%%%%%%%%%%%%%%%%%%%%%%%%%%%%%%%%%%%%%%%%%%
\begin{equation}
\Delta E=6B+5C-\Delta,
\label{Eq:RacPrm}
\end{equation}
%%%%%%%%%%%%%%%%%%%%%%%%%%%%%%%%%%%%%%%%%%%%%%%%%%
where $B$ and $C$ are the Racah parameters. Using the values for free Mn$^{3+}$ ion, $B_0$=120 meV and $C_0$=552 meV, we can obtain the orbital reduction factor $x$ from Eq.~(\ref{Eq:RacPrm}) by setting $B=xB_0$ and $C=xC_0$. Then, the value of $\lambda$ in the $R$MnO$_3$ compounds can be obtained from $\lambda=x\lambda_0$ with $\lambda_0$=10.5 meV being the coupling constant for free Mn$^{3+}$ ion.

%%%%%%%%%%%%%%%%%%%%%%%%%%%%%%%%%%%%%%%%%%%%%%%%%%%%%%%%%%%%%%%%%%
\begin{table}
\caption{Energy levels $\Delta E$ and $\Delta$ for Mn$^{3+}$ ion surrounded by octahedrally coordinated oxygens obtained in the optical absorption measurements from Refs.~\cite{Matsumoto70,Gerristen63}. Calculated orbital reduction factor $x$ and coupling constant of the spin-orbit interaction $\lambda$ are also listed.}
\begin{tabular}{ccccc}
\hline
$\Delta E$ (meV) & $\Delta$ (meV) & $x$ & $\lambda$ (meV) & Ref \\
\hline
353.4 & 2530.8 & 0.83 & 8.75 & \cite{Matsumoto70} \\ 
349.7 & 2083.2 & 0.71 & 7.38 & \cite{Gerristen63} \\
\hline
\end{tabular}
\label{tabl:Rachprm}
\end{table}
%%%%%%%%%%%%%%%%%%%%%%%%%%%%%%%%%%%%%%%%%%%%%%%%%%%%%%%%%%%%%%%%%%
The values of $\Delta E$ and $\Delta$ for Mn$^{3+}$ ion surrounded by octahedrally coordinated oxygens are experimentally obtained in the optical absorption spectra~\cite{Matsumoto70,Gerristen63}. In Table.~\ref{tabl:Rachprm}, we list the values of $\Delta E$ and $\Delta$ taken from two different references as well as the values of $x$ and $\lambda$. Using these parameter values and structural data in Refs.~\cite{RodriguezC98,Alonso00,Mori02,Dabrowski05,Tachibana07}, we calculate $D$ and $E$ for $R$MnO$_3$ with various $R$ ions, which are plotted in Figs.~\ref{Fig19}(a) and \ref{Fig19}(b) as functions of the ionic $R$-site radius. Note that the difference of $\Delta E$ and $\Delta$ between two references causes a small difference of calculated values of $D$ and $E$. In our model calculation, we use the values obtained from the optical data in Ref.~\cite{Gerristen63}, because with these values, the calculation reproduces the experimental phase diagrams with a reasonable set of the DM parameters, which is consistent with the band calculation~\cite{Solovyev96} and the ESR experiments~\cite{Tovar99,Deisenhofer02} as discussed in Sec.~\ref{SSec:Model}. On the other hand, with the values obtained from the optical data in Ref.~\cite{Matsumoto70}, we need to assume rather large DM vectors on the out-of-plane bonds to reproduce the experimental phase diagrams.

%%%%%%

\end{document}